\documentclass[journal]{IEEEtran}
\ifCLASSINFOpdf

\else
  
\fi

\usepackage{cite}
\usepackage{amsfonts}
\usepackage{amsmath}
\usepackage[pdftex]{graphicx}
\usepackage{amsmath,amssymb,amsfonts}
\usepackage{graphicx}
\usepackage{booktabs} %表格线条加粗所需
\usepackage{textcomp}
\usepackage{xcolor}
\usepackage{float}
\usepackage{amsmath, bm}
\usepackage[ruled,linesnumbered,vlined]{algorithm2e}

\usepackage{stfloats}

\usepackage{CJK}
\usepackage{indentfirst}
\usepackage{amsmath}
\usepackage{cases}

\usepackage{hyperref}
\hypersetup{
            colorlinks=true,
            linkcolor=blue,
            anchorcolor=blue,
            citecolor=blue,
            filecolor=blue,
            urlcolor=blue
            }
\makeatletter
\newcommand{\ccc}[1]{{\color{blue}\cite{#1}}}
\makeatother

\definecolor{b}{rgb}{0,0,0}

\usepackage{amsmath}

\begin{document}

\title{Double-Edge-Assisted Computation Offloading and Resource Allocation for Space-Air-Marine Integrated Networks}

 \author{Zhen~Wang,~\IEEEmembership{Student Member,~IEEE,}
         Bin~Lin,~\IEEEmembership{Senior Member,~IEEE,}
        and~Qiang (John) Ye,~\IEEEmembership{Senior Member,~IEEE}
%         Yuguang~Fang,~\IEEEmembership{Fellow,~IEEE}
%         and~Xiaoling~Han,~\IEEEmembership{Student Member,~IEEE}% <-this % stops a space
\thanks{Copyright (c) 20xx IEEE. Personal use of this material is permitted. However, permission to use this material for any other purposes must be obtained from the IEEE by sending a request to pubs-permissions@ieee.org.}
\thanks{\emph{(Corresponding author: Bin Lin.) }}
\thanks{Zhen Wang is with the Information Science and Technology College, Dalian Maritime University, Dalian, 116026, China, and Communication Engineering of Dalian Neusoft University of Information, Dalian, 116023, China. E-mail: wangzhen\underline{~}jsj@neusoft.edu.cn.}% <-this % stops a space
\thanks{Bin Lin is with the Information Science and Technology College, Dalian Maritime University, Dalian, 116026, China. E-mail: binlin@dlmu.edu.cn.}% <-this % stops a space
\thanks{Qiang (John) Ye is with the Department of Electrical and Software Engineering, Schulich School of Engineering, University of Calgary, 2500 University Drive NW, Calgary, AB T2N 1N4. Email: qiang.ye@ucalgary.ca.}% <-this % stops a space
\thanks{This manuscript has been accepted by  IEEE Transactions on Vehicular Technology, DOI: 10.1109/TVT.2025.3561346.}% <-this % stops a space

 }

\maketitle

\begin{abstract}
%Space-air-marine integrated network (SAMIN) is an emerging maritime network architecture devised to facilitate seamless and efficient wireless communication services for marine applications.  
%To tackle the immense demand for computation-intensive and delay-sensitive marine applications and services within SAMIN, Multi-access Edge Computing (MEC) is envisioned as a promising solution to provide robust computing capabilities for resource-constrained marine devices. 
In this paper, we propose a double-edge-assisted computation offloading and resource allocation scheme tailored for space-air-marine
integrated networks (SAMINs). Specifically, we consider a scenario where both unmanned aerial vehicles (UAVs) and a low earth orbit (LEO) satellite are equipped with edge servers, providing computing services for maritime autonomous surface ships (MASSs). Partial computation workloads of MASSs can be offloaded to both UAVs and the LEO satellite, concurrently, for processing via a multi-access approach.
To minimize the energy consumption of SAMINs under latency constraints, we formulate an optimization problem and propose energy efficient algorithms to jointly optimize offloading mode, offloading volume, and computing resource allocation of the LEO satellite and the UAVs, respectively. We further exploit an alternating optimization (AO) method and a layered approach to decompose the original problem to attain the optimal solutions.
Finally, we conduct simulations to validate the effectiveness and efficiency of the proposed scheme in comparison with benchmark algorithms.

\end{abstract}

\begin{IEEEkeywords}
Space-air-marine integrated networks, 6G, maritime multi-access edge computing, double-edge-assisted computation offloading, offloading mode and volume, computing resource allocation.
\end{IEEEkeywords}

\IEEEpeerreviewmaketitle

\section{Introduction}

\IEEEPARstart{W}{ith} the unprecedented development of maritime activities (e.g., marine resource exploration, object recognition, and intelligence reconnaissance), a significant proliferation of marine wireless devices is underway to gather immense amounts of oceanic data for diverse maritime services\ccc{10196037},\ccc{9800125},\ccc{10018447}.  
\textcolor{b}{For instance, in the context of marine environmental monitoring and real-time data processing, maritime autonomous surface ships (MASSs) are equipped with a variety of sensors, including cameras, Light Detection and Ranging (LiDAR), millimeter-wave radar, inertial measurement units (IMUs), and global positioning systems (GPS),  which enable the MASSs to collect multidimensional data on weather conditions, water quality, and marine biological activities in real-time.} In marine disaster relief, MASSs are employed to capture images and videos of search and rescue scenes to verify the targets and subsequently enhance the overall efficiency and effectiveness of the rescue endeavors\ccc{10490255},\ccc{10499870}. However, the scarcity of conventional maritime communication and computing resources poses a significant challenge in fulfilling the stringent requirements of such high-reliability and low-latency applications\ccc{9800919}. To mitigate the impediment, it is imperative to conduct more efficient communication and computing in maritime networks, which has garnered substantial interest from both academia and industry in recent years.
  
Multi-access edge computing (MEC) has emerged as a highly effective approach, significantly enhancing computing efficiency and minimizing decision-making latency for resource-constrained marine devices\ccc{10757774},\ccc{10376220},\ccc{9454395},\ccc{9877926}. 
\textcolor{b}{Leveraging MEC, the MASSs are able to rapidly offload and process large volumes of sensor data and generate real-time environmental insights. Moreover, the MASSs can also perform path planning and autonomous navigation based on the environmental data collected in real time.} 
Recently, significant research efforts have been put towards providing innovative methodologies for maritime MEC to bolster the performance and efficiency of marine networks.
In\ccc{9881915}, Li \emph{et al.} focused on the applications of unmanned aerial vehicles (UAVs) for autonomous detection and tracking in a maritime environment and proposed a task offloading scheme to minimize the system energy consumption. 
In\ccc{10449425}, Zeng \emph{et al.} introduced an energy-efficient collaborative computation offloading scheme utilizing unmanned surface vehicle (USV) fleets to support smart maritime services, where UAVs act as service requesters and USV fleets serve as helpers facilitating the computation offloading process.
In the paradigm of MEC, the offloading and computing efficiency can be improved by segmenting computation loads into multiple parts which are then offloaded to different edge servers for further processing\ccc{10601175},\ccc{8419289}. \textcolor{b}{This distributed approach can improve the system performance, through parallel processing, and cost efficiency by sharing resources, making it ideal for big data, real-time applications, and global systems\ccc{10026506},\ccc{10841367}.}
% On this basis, how to design efficient maritime MEC schemes  should be further exploited. 

The evolution of the six generation (6G) wireless technologies is driving the integration of multidimensional wireless communication resources, encompassing space, air, sea, and ground networks, to achieve ubiquitous communication coverages\ccc{10571560},\ccc{8344453},\ccc{10323351},\ccc{10700788}. In this context, a space-air-marine integrated network (SAMIN) emerges to proficiently harness diverse resources to empower intelligent network control and efficient wireless communication services for marine applications. Lin \emph{et al.} proposed a space-air-ground-sea integrated network architecture and jointly optimized the offloading strategies and resource allocation to minimize the energy consumption of the whole system in\ccc{10518032}. 
Wang \emph{et al.} proposed a double-edge secure offloading scheme for SAMINs, where computing workloads can be processed on both base stations (BSs) and satellites for delay-sensitive applications in\ccc{10092853}.
The MEC paradigm exhibits immense potential in supporting various services facilitated by satellite-assisted networks, for addressing the computation-intensive and delay-sensitive service requirements in the oceanic realm. Given the energy and computing constraints of a single edge node, a viable solution to bolster edge computing efficiency involves distributing oceanic task computing workloads simultaneously among space, air, and marine devices for parallel processing. 
%To the best of our knowledge, how to optimize the computation offloading and resource allocation for MASSs with respect to the double-edge-assisted SAMIN architecture has not been fully exploited. 

\textcolor{b}{The combination of UAVs and MEC has been studied to enhance edge computing performance in marine environments\ccc{10490255},\ccc{10376220},\ccc{10224351},\ccc{10778110},\ccc{9994654}. 
Considering the typical constraints and dynamics in communication, computation, and energy resources associated with a single UAV edge, a double-edge-assisted SAMIN architecture can make better utilization of various resources to improve computing efficiency for marine devices. 
The ``double-edge'' emphasizes on the collaborative and hierarchical nature of the two edge computing layers, i.e., the UAVs as the first edge layer and the low earth orbit (LEO) satellite as the second edge layer. The UAVs provide computation resources in close proximity to the MASSs, which is particularly effective in handling time-sensitive tasks due to low-latency communication and efficient task offloading. The LEO satellites act as the second edge computing layer, offering broader coverage and significant computational capabilities, which is suitable for handling computationally intensive tasks and provides backup when UAVs are unavailable or overloaded. 
By leveraging the complementary strengths of UAVs (proximity and low latency) and LEO satellites (extensive coverage and powerful computation), the double-edge-assisted SAMIN provides a robust and flexible solution for task offloading in maritime environments. This dual-layer approach ensures that MASSs can offload tasks efficiently, even in dynamic and challenging conditions, such as varying channel quality, mobility, and resource availability.
However, there are also key technical challenges to overcome under this layered edge computing architecture: 1) where to offload tasks for processing; 2) how to assign tasks to different edge servers; 3) how computing and communication resources are allocated among edge computing nodes to facilitate efficient task processing for MASSs.}

\textcolor{b}{In this paper, we propose a double-edge-assisted MEC system for an SAMIN, where an LEO satellite and UAVs are equipped with edge computing resources, enabling them to concurrently provide computational services for marine devices. 
The computation workloads of marine devices can be offloaded to the LEO satellite and UAVs simultaneously via a multi-access approach. 
To our knowledge, no pertinent research exists on employing UAVs and satellites as double-edge servers to furnish edge computing services for MASSs at the same time.
The key contributions of this paper mainly include the following aspects:}

\begin{itemize}\color{b}
\item[$\bullet$] 
\emph{Double-edge-assisted Task Offloading Framework}: We propose a novel double-edge-assisted computation offloading framework for an SAMIN. In this framework, both the LEO satellite and UAVs serve as BSs equipped with edge servers to provide computing services for MASSs. The tasks generated by the MASSs are divisible and can be offloaded in parallel to the LEO satellite and UAVs for processing through a multi-access approach. This dual-layer architecture leverages the complementary strengths of UAVs and the LEO satellite, enabling efficient and flexible task offloading in dynamic maritime environments.
\item[$\bullet$] 
\emph{Optimization and Implementation Methodologies}: To capitalize on the heterogeneous computing resources provided by the UAVs and the LEO satellite, we propose a joint computation offloading and resource allocation scheme to enhance the communication and computing efficiency with the objective of minimizing the energy consumption of the SAMIN. 
We employ the alternating optimization (OA) method and propose a layered approach to solve the complex optimization problem. Specifically, we jointly optimize the offloading mode, the offloading volume, and the computing resource allocation of both the LEO satellite and UAVs, respectively, to attain an efficient and scalable solution that adapts to the dynamic nature of the SAMIN.
\item[$\bullet$] 
\emph{Performance Evaluation}: We perform extensive numerical analysis to validate the efficacy of the proposed computation offloading and resource allocation scheme. The numerical results demonstrate that the proposed algorithms significantly minimize the energy dissipation of the SAMIN and confirm the effectiveness and efficiency of the algorithms when compared to the state-of-the-art.
\end{itemize}

The remainder of this paper is organized as follows. Section II presents the related work. Section III describes the system model of the SAMIN under consideration. The problem formulation and the energy-efficient double-edge-assisted task offloading framework are presented in Sections IV and V, respectively. 
Section VI presents the performance evaluation, and Section VII draws concluding remarks and discusses future research directions.
% Table I lists the key notations according to the order in which they first appear in the paper.

% \begin{table}[htbp]
% \centering
% \caption{List of Key Notations}
% \begin{tabular}{c|c}%l=left, r=right,c=center分别代表左对齐，右对齐和居中，字母的个数代表列数
% \hline
% \hline
% $\mathbf{Symbol}$ & $\mathbf{Meaning}$  \\ 
% \hline
% $M_k$  &The numbedr of TUs under MIS $k$ \\
% \hline
% $p_{k,i}(t)$  &Processing power of MIS $k$ for TU $i$\\
% \hline
% \hline
% \end{tabular}
% \end{table}

\section{Related Work}
\subsection{Terrestrial/Air-Assisted MEC in Maritime Networks}
Existing studies have developed approaches for terrestrial/air-assisted maritime MEC to enhance the computing efficiency of marine applications and services. Current research endeavors can be mainly categorized into two representative approaches to facilitate computation offloading for marine devices. One approach focused on an offshore scenario where the computing workloads of marine devices are directly offloaded or relayed to coastal BSs for further processing\ccc{10224351},\ccc{10453323},\ccc{9693477}. 
Dai \emph{et al.} proposed a UAV-assisted data offloading scheme where each UAV served as a relay node for smart containers to transfer workloads to coastal BSs in\ccc{9693477}.
In\ccc{10453323}, Wang \emph{et al.} established an MEC-enabled sea lane monitoring network (MSLMN) architecture where tasks can be offloaded to buoys or coastal base stations for processing.
The second approach explored the utilization of UAVs\ccc{10118952}, high altitude platforms (HAPs)\ccc{9996964}, sea surface-stations\ccc{9978925}, floating platforms\ccc{10173733} or satellites as edge computing platforms to carry out task computation. 
In\ccc{9996964}, Li \emph{et al.} exploited a HAP to perform computation offloading and provide cooperative jamming for the communication security of USVs.
In\ccc{10499911}, Dai \emph{et al.} proposed a multi-UAV facilitated MEC framework tailored specifically for marine networks, aiming to optimize both operational efficiency and resource utilization.
%To the best of our knowledge, there exists no current research that explores the utilization of both UAVs and satellites as edge servers simultaneously for processing maritime workloads in parallel. 

Considering multiple tiers of terrestrial/air network platforms, how to efficiently utilize the communication and computing resources of different network tiers is the key to \textcolor{b}{improving} the processing performance of maritime computing workloads.

\subsection{SAMIN-Assisted Communication and Computing}

The 6G wireless networks are poised to transcend geographical boundaries, achieving seamless global coverage and effectively alleviating traffic congestion with the integration of satellite networks\ccc{10409745}. As an enhancement to conventional terrestrial networks, SAMINs can provide reliable network connectivity and distributed computing resources from marine edge nodes\ccc{9941481},\ccc{10817534},\ccc{10040542}.
Li \emph{et al.}\ccc{9979795} proposed a space-air-ground-ocean-integrated network (SAGOI-Net) framework, where an intelligent autonomous underwater glider (AUG) is employed to serve marine applications. 
Jung \emph{et al.} proposed an innovative hybrid approach that integrates LEO satellite with UAVs to provide computing services in space-air-sea integrated networks for marine Internet-of-Things (IoT) systems\ccc{10155303}. 
Guo \emph{et al.}\ccc{10254489} proposed a bidirectional multibeam transmit reflect array (TRA) antenna for a space-air-ground-sea integrated network (SAGSIN) to facilitate tri-beam transmission and dual-beam reflection. 
To provide good quality-of-service (QoS) for SAGOI-Net, Zhang \emph{et al.} proposed a multi-domain virtual network embedding solution\ccc{10414013}, while Lin \emph{et al.} proposed two resource management schemes based on deep reinforcement learning (DRL) to satisfying QoS requirements for SAGSINs\ccc{10660307}.

Although most existing studies focus on MEC-enabled marine networks and SAMIN-assisted communication and computing, several crucial aspects remain unexplored: 1) limited resource availability on marine edge servers; 2) integrating satellites with UAVs to provide double-edge-assisted computing services to marine devices; 3) joint optimization of computation offloading and resource allocation within a double-edge-assisted SAMIN architecture. 

\section{System model}
This section first introduces a double-edge-assisted SAMIN, where MASSs can offload their tasks to UAVs and an LEO satellite edge simultaneously. Then, the communication, computing, and task offloading models are presented.
\subsection{Network model}
We consider an SAMIN consisting of one LEO satellite and multiple UAVs, cooperatively supporting task offloading for connected MASSs, as shown in Fig. 1. 
\begin{figure}[htbp]
\centerline{\includegraphics[width=0.6\textwidth]{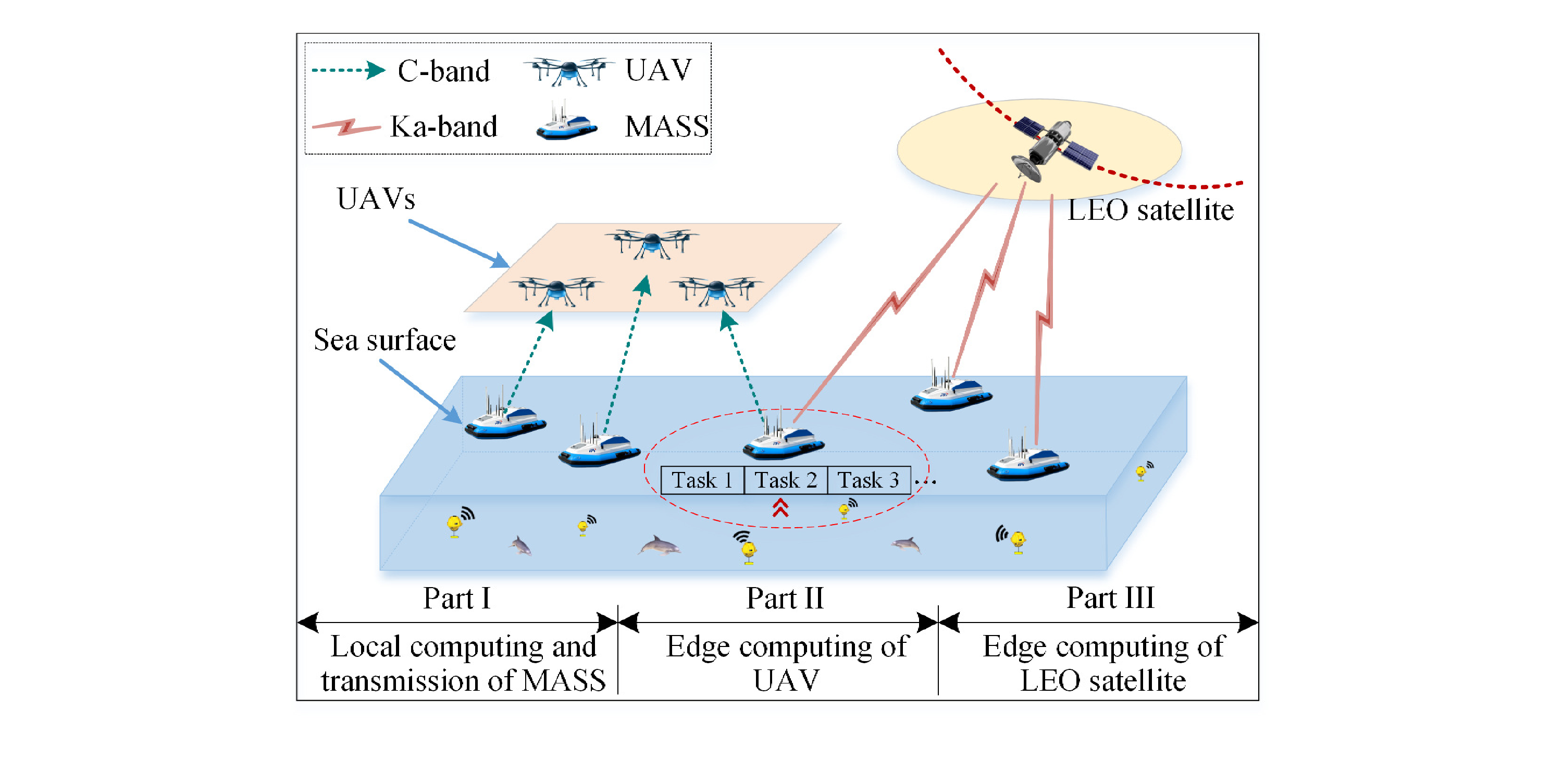}}
\caption{Network model.}
\label{fig}
\end{figure}
In the considered scenario, an LEO satellite with edge computing capacity coordinates with each UAV to assist the task processing of MASSs.
A group of MASSs are distributed on the sea surface to monitor the marine environment and collect oceanic data (e.g., aquaculture monitoring videos, real-time data sensing). Meanwhile, a certain number of UAVs equipped with computing capacities are dispatched to receive critical data collected by MASSs. We assume each UAV serves $N$ MASSs, and the set of UAVs and the set of MASSs under each UAV are denoted as $\mathcal{M}=\left\{1, 2,..., m,..., M\right\}$ and $\mathcal{N}=\left\{1, 2,..., n,..., N\right\}$, respectively. For convenience, we denote the $m$-th UAV and its connected $n$-th MASS as $U_m$ and $M_{mn}$, respectively, where we have $m\in\mathcal{M}$ and $n\in\mathcal{N}$. The input task data size (in bits) of $M_{mn}$ is denoted as $S_{mn}$, with $s_{mn}$ indicating the number of bits intended for offloading, and the task is processed only locally when $s_{mn}=0$. We denote $a_{mn}\in[0,1]$ as the offloading ratio of $s_{mn}$ from $M_{mn}$ to $U_m$. Then, the task workloads of $M_{mn}$ offloaded to a LEO satellite is denoted as $(1-a_{mn})s_{mn}$.

Specifically, the overall execution process of double-edge-assisted computing for one task comprises three parts: 1) One MASS executes local task computing and uploads its partial workloads (if any) to one UAV and/or the LEO satellite for further computation; 2) The UAV receives and processes the offloaded data from the MASS; 3) The LEO satellite receives the offloaded data from the MASS for processing. 

The orthogonal frequency-division multiple access (OFDMA) protocol is employed for each UAV or the LEO satellite channel access. Specifically, different UAVs reuse the same portion of spectrum resources within C-band, which are then equally divided and allocated to MASSs. With OFDMA and the assumption of non-overlapping UAV coverages, the intra-cell interference among MASSs under each UAV and the inter-cell interference among UAVs are assumed to be negligible.

We utilize \textcolor{b}{a three-dimensional (3D) Cartesian coordinate} to delineate the positions of UAVs and MASSs. The proximity between MASSs and UAVs affects the channel link quality. Note that we assume the locations of MASSs and UAVs remain stable during the data transmission. Let $\textbf{q}_m=[x_m,y_m,z_m]^T\in \mathbb{R}^{3\times1}$ denote the spatial coordinates of $U_m$, in which $x_m$, $y_m$, and $z_m$ correspond to the longitude, latitude, and height of $U_m$, respectively. Let $\textbf{q}_{mn}=[x_{mn},y_{mn},z_{mn}]^T\in \mathbb{R}^{3\times1}$
denote the spatial coordinates of $M_{mn}$.

To ensure the wireless channel quality during the data transmission, we impose a constraint that the distance between $U_m$ and $M_{mn}$ must not surpass the maximum allowable communication distance $d^{\max}$, which is expressed as
\begin{equation}
\vert\vert\textbf{q}_m-\textbf{q}_{mn}\vert\vert\leq d^{\max}.
\end{equation}

\subsection{Communication model}

\emph{(1) Communication Model from MASSs to UAVs}

\textcolor{b}{
We assume that the altitude of the UAVs is sufficient for line-of-sight (LoS) transmission. Considering the uniqueness of the marine environment, e.g., the strong direct signal, the primary factors affecting the overseas wireless channel are multi-path effects caused by ocean waves and extreme weather conditions. We model the communication between an MASS and a UAV as an air-sea channel exhibiting Rician fading\ccc{8960465}\ccc{7407385}, which is considered a combination of large-scale and small-scale fading, as explained below.}

\textcolor{b}{ 
The large-scale path loss model is expressed as
\begin{equation}
	L_{mn}^U(dB)=L_0+10\zeta\log_{10}\left(\frac{d_{mn}^U}{d_0} \right)+X_{\sigma_X}+\xi F ,
\end{equation}
where $d_{mn}^U=\sqrt{(x_m-x_{mn})^2+(y_m-y_{mn})^2+(z_m-z_{mn})^2}$ denotes the distance between $U_m$ and $M_{mn}$, $L_0$ is the path loss at the reference distance $d_0$, $\zeta$ indicates the path-loss exponent due to the ducting effect over the sea surface\ccc{8528349}, $X_{\sigma_X}\in \mathcal{CN}(0,\sigma_X)$ denotes the shadow fading caused by, e.g., sea waves under high sea state conditions, $F$ is an adjustment parameter for direction of travel, and $\xi$ is set to $1$ or $-1$ to indicate the moving direction of the UAVs (towards or away from the ground site)\ccc{8528349}.}

\textcolor{b}{
The small-scale Rician fading $\tilde{\Lambda}_{mn}^U$ is represented as
\begin{equation}
	\tilde{\Lambda}_{mn}^U=\sqrt{\frac{K_0}{1+K_0}}+\sqrt{\frac{1}{1+K_0}}o_{mn}^U,
\end{equation}
where $o_{mn}^U \in \mathcal{CN}(0,1)$ and $K_0$ is the Rician factor.}
\textcolor{b}{
Then, the channel coefficient is formulated as
\begin{equation}
	G_{mn}^U=\left(L_{mn}^U\right)^{-1/2} \tilde{\Lambda}_{mn}^U.
\end{equation}
}
\textcolor{b}{
The channel gain between $U_m$ and $M_{mn}$ is expressed as
\begin{equation}
	g_{mn}^U=G^UG^M\mid G_{mn}^U\mid^2
\end{equation}
where $G^U$ and $G^M$ are the antenna gain of the UAVs and MASSs, respectively.}
According to the Shannon capacity theorem, the transmission rate (link capacity) $R_{mn}^{U}$ between $U_m$ and $M_{mn}$ is calculated as
\begin{equation}
R_{mn}^U=W_{mn}^U\log_2\left(1+\frac{p_{mn}^Ug_{mn}^U}{\sigma^2}\right) 
\end{equation}
where $W_{mn}^U$ denotes the transmission bandwidth of $M_{mn}$ and $\sigma^2$ is the spectral power of the additive white Gaussian noise (AWGN). 
Let $t_{mn}^U$ denote the transmission time for offloading partial workloads $a_{mn}s_{mn}$ from $M_{mn}$ to $U_m$, satisfying
\begin{equation}
t_{mn}^U=\frac{a_{mn}s_{mn}}{R_{mn}^U}.
\end{equation}
Based on (3) and (4), we obtain the required transmission power of $M_{mn}$ for offloading workloads $a_{mn}s_{mn}$ to $U_m$ as 
\begin{equation}
p_{mn}^U=\frac{\sigma^2}{g_{mn}^U}\left(2^{\frac{a_{mn}s_{mn}}{t_{mn}^UW_{mn}^U}}-1\right).
\end{equation}
Then, the corresponding energy consumption is expressed as
\begin{equation}
e_{mn}^U=p_{mn}^Ut_{mn}^U=\frac{t_{mn}^U\sigma^2}{g_{mn}^U}\left(2^{\frac{a_{mn}s_{mn}}{t_{mn}^UW_{mn}^U}}-1\right).
\end{equation}

\emph{(2) Communication Model from MASSs to the LEO satellite}

\emph{(i) Coverage Time Model of the LEO satellite:} 
Different from the terrestrial MEC network model, the location of the LEO satellite changes dynamically. Hence, an MASS cannot always communicate with the LEO satellite at any time. We obtain the maximum communication time between $M_{mn}$ and LEO satellite, denoted as
\begin{equation}
T^{\max}=\frac{2(R_e+h)\cdot \phi_{mn}^L}{v^L}
\end{equation}
where $v^L=\sqrt{K_0/(R_e+h)}$ is the moving speed of the LEO satellite, $h$ represents the height of the LEO satellite orbit,  $R_e$ denotes the radius of the earth, $\phi_{mn}^L$ is the geocentric angle between $M_{mn}$ and the LEO satellite, which is expressed as 
\begin{equation}
\phi_{mn}^L=\arccos\left(\frac{R_e}{R_e+h}\cdot \cos\theta_{mn}^L\right)-\theta_{mn}^L.
\end{equation}
In (8), $\theta_{mn}^L$ is the elevation angle between $M_{mn}$ and the LEO satellite.
Considering the LEO satellite is with high moving speed, the communication between ground users and the LEO satellite is limited by the user coverage time of the LEO satellite.

\emph{(ii) Communication Model from MASSs to the LEO satellite:}
For the LEO satellite communications, we assume that the position information of the LEO satellite is known to all MASSs due to the orbital pre-planning within one time slot. For simplicity, we consider a quasi-static fading channel model. Then, the transmission rate of $M_{mn}$ is formulated as 
\begin{equation}
R_{mn}^L=W_{mn}^L\log_2\left(1+\frac{p_{mn}^L\mid h_{mn}^L\mid^2}{W_{mn}^LN_0}\right).
\end{equation}
In (9), we have $ h_{mn}^L=g_{mn}^L\cdot\beta_{mn}^L\cdot(d_{mn}^L)^{-\gamma/2}$, where $g_{mn}^L$ is a complex Gaussian variable representing Rayleigh fading, $\beta_{mn}^L$ denotes the fading involving shadowing, rain, and other fading, $\gamma$ is the path exponent, and $d_{mn}^L=\sqrt{R_e^2+(R_e+h)^2-2R_e(R_e+h)\cos\phi_{mn}^L}$ represents the distance from $M_{mn}$ to the LEO satellite.

However, the distance between $M_{mn}$ and the LEO satellite is relatively long, causing $M_{mn}$ to be affected by the propagation delay when communicating with the LEO satellite. Thus, the transmission time $t_{mn}^L$ consists of the propagation delay and the transmission delay, which is expressed as 
\begin{equation}
t_{mn}^L=\frac{(1-a_{mn})s_{mn}}{R_{mn}^{L}}+\frac{2d_{mn}^{L}}{c}
\end{equation}
where $c$ denotes the speed of light.
Based on (9) and (10), we obtain the required transmission power of $M_{mn}$ for offloading partial workloads to the LEO satellite as
\begin{equation}
p_{mn}^L=\frac{W_{mn}^LN_0}{\mid h_{mn}^L\mid^2}\left[2^{\frac{(1-a_{mn})s_{mn}}{\left(t_{mn}^L-\frac{2d_{mn}^{L}}{c}\right)W_{mn}^L}}-1\right].
\end{equation}
The corresponding energy consumption is formulated as
\begin{equation}
e_{mn}^L=p_{mn}^Lt_{mn}^L=\frac{t_{mn}^LW_{mn}^LN_0}{\mid h_{mn}^L\mid^2}\left[2^{\frac{(1-a_{mn})s_{mn}}{\left(t_{mn}^L-\frac{2d_{mn}^{L}}{c}\right)W_{mn}^L}}-1\right].
\end{equation}

\subsection{Computation Model}

\emph{(1) Local computing at MASSs: }

Considering the case that the task of $M_{mn}$ is partially processed locally, we denote $\rho_{mn}^l$ as the CPU computing capacity of $M_{mn}$, which is quantified by the number of CPU cycles per second. Then, the execution time of local computing for $M_{mn}$ is given by
\begin{equation}
T_{mn}^l=\frac{(S_{mn}-s_{mn})c_{mn}}{\rho_{mn}^l}
\end{equation}
where $c_{mn}$ denotes the number of CPU cycles for processing one bit of data by $M_{mn}$.
The corresponding energy consumption $E_{mn}^l$ is computed as
\begin{equation}
E_{mn}^l=P_{mn}^lT_{mn}^l=P_{mn}^l\frac{(S_{mn}-s_{mn})c_{mn}}{\rho_{mn}^l}
\end{equation}
where $P_{mn}^l$ is the power consumption of $M_{mn}$ for local computing.

\emph{(2) Edge computing at UAVs: }

When partial task of $M_{mn}$ is offloaded to $U_m$, we denote $\rho_{mn}^{U}$ and $\rho_m^{\max}$ as the computation capacity of $U_m$ allocated to $M_{mn}$ and the maximum number of executable CPU cycles at $U_m$, respectively, satisfying $\sum_{n\in\mathcal{N}}\rho_{mn}^{U}\leq\rho_m^{\max}$.
Then, the processing latency of $U_m$ to complete the assigned workloads is denoted as
\begin{equation}
T_{mn}^U=\frac{a_{mn}s_{mn}c_m}{\rho_{mn}^{U}}
\end{equation}
where $c_m$ denotes the number of CPU cycles for processing one bit of data by $U_m$.

The corresponding energy consumption $E_{mn}^U$ is computed as:
\begin{equation}
E_{mn}^U=P_{mn}^UT_{mn}^U=P_{mn}^U\frac{a_{mn}s_{mn}c_m}{\rho_{mn}^{U}}
\end{equation}
where $P_{mn}^U$ is the power consumption of $U_m$ for edge computing.

\emph{(3) Edge computing at the LEO satellite: }

When partial task of $M_{mn}$ is offloaded to the LEO satellite, we denote $\rho_{mn}^{L}$ and $\rho_{\max}^L$ as the computational capacity of the LEO satellite allocated to $M_{mn}$ and the maximum number of executable CPU cycles at LEO satellite, respectively, satisfying $\sum_{m\in\mathcal{M}}\sum_{n\in\mathcal{N}}\rho_{mn}^{L}\leq\rho_{\max}^L$.
Then, the processing latency at the LEO satellite to complete the assigned workloads is denoted as
\begin{equation}
T_{mn}^L=\frac{(1-a_{mn})s_{mn}c^L}{\rho_{mn}^{L}}
\end{equation}
where $c^L$ denotes the number of CPU cycles for processing one bit of data by the LEO satellite.
The corresponding energy consumption is calculated as
\begin{equation}
E_{mn}^L=P_{mn}^LT_{mn}^L=P_{mn}^L\frac{(1-a_{mn})s_{mn}c^L}{\rho_{mn}^{L}}
\end{equation}
where $P_{mn}^L$ is power consumption of the LEO satellite for edge computing.

The overall latency and energy consumption associated with completing $M_{mn}$’s workloads is denoted as 
\begin{equation}
T_{mn}^{tot}=\max\left\{T_{mn}^l,t_{mn}^U+T_{mn}^U,t_{mn}^L+T_{mn}^L\right\}
\end{equation}
and
\begin{align}
E_{mn}^{tot}&=E_{mn}^l+e_{mn}^U+e_{mn}^L+E_{mn}^U+E_{mn}^L\nonumber\\
&=P_{mn}^l\frac{(S_{mn}-s_{mn})c_{mn}}{\rho_{mn}^l}+P_{mn}^U\frac{a_{mn}s_{mn}c_m}{\rho_{mn}^{U}}\nonumber\\
&+\frac{t_{mn}^U\sigma^2}{g_{mn}^U}\left(2^{\frac{a_{mn}s_{mn}}{t_{mn}^UW_{mn}^U}}-1\right)+P_{mn}^L\frac{(1-a_{mn})s_{mn}c^L}{\rho_{mn}^{L}}\nonumber\\
&+\frac{t_{mn}^LW_{mn}^LN_0}{\mid h_{mn}^L\mid^2}\left[2^{\frac{(1-a_{mn})s_{mn}}{\left(t_{mn}^L-\frac{2d_{mn}^{L}}{c}\right)W_{mn}^L}}-1\right],
\end{align}
respectively.
Thus, the overall energy dissipation of the proposed system is formulated as
\begin{align}
E^{tot}&=\sum_{m\in\mathcal{M}}\sum_{n\in\mathcal{N}}E_{mn}^{tot}.
\end{align}

\subsection{Offloading Model}
Due to the concurrent availability of both UAVs and the LEO satellite, it is flexible for an MASS to offload its task to the LEO satellite and UAVs for processing whenever its computation capacity is insufficient to fulfill the task processing demands. It is crucial to note that the optimization of the offloading policy is undertaken at the LEO satellite level.
To fulfill the computation objectives of MASSs, we present an offloading model comprising the following four phases, as illustrated in Fig. 2. 
\begin{figure}[htbp]\color{b}
\centerline{\includegraphics[width=0.45\textwidth]{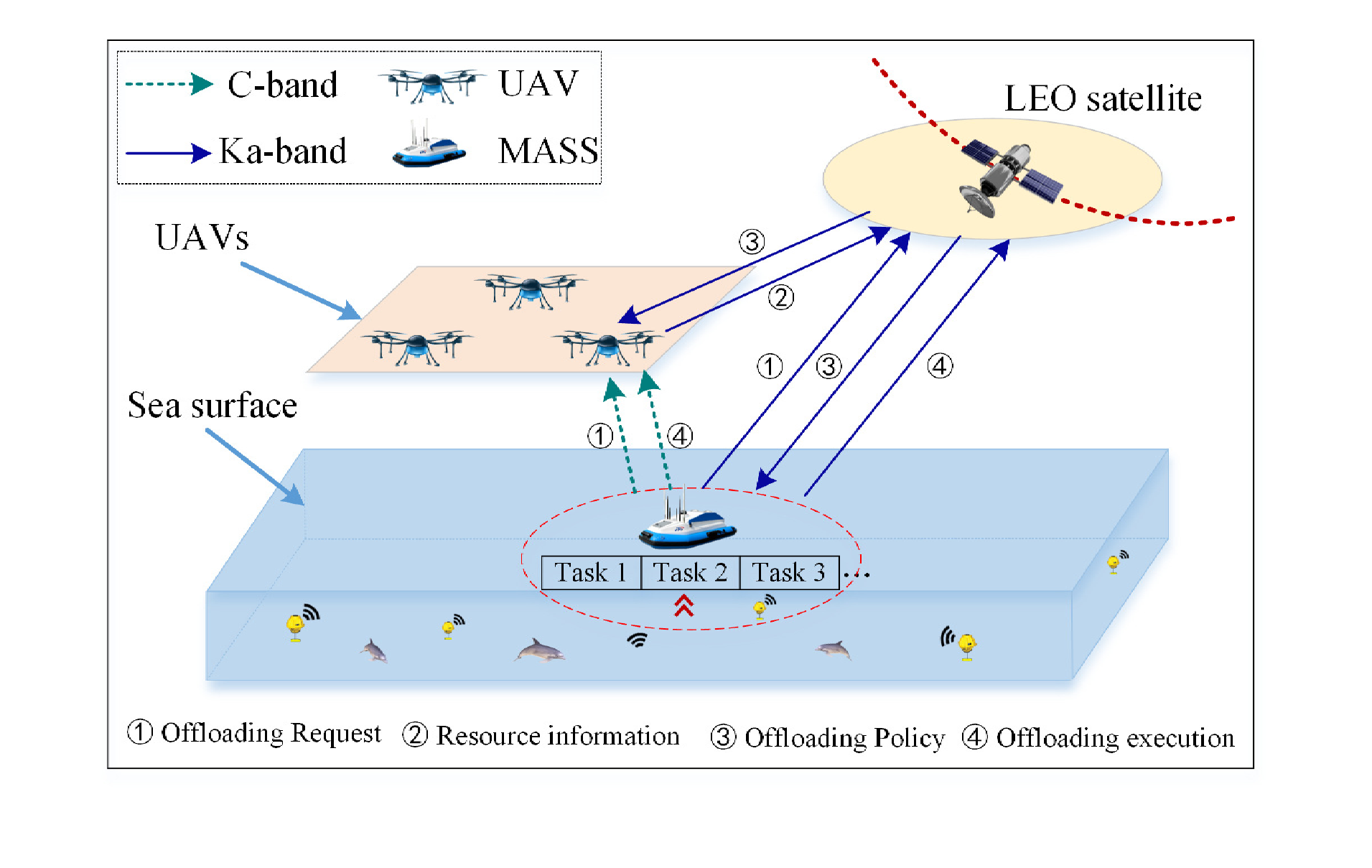}}
\caption{Offloading process.}
\label{fig}
\end{figure}

\emph{(1) Offloading Request: }At the beginning of each time slot, $M_{mn}$ sends its offloading request to its serving $U_m$ and the LEO satellite over the C-band and/or the Ka-band, respectively.

\emph{(2) Resource Information Notification: }Upon receiving the request message, $U_m$ promptly reports its resource status to the LEO satellite.

\emph{(3) Offloading Policy: }Upon receiving the request message and resource status of UAVs, the LEO satellite devises an offloading policy aiming at allocating the requested communication and computing resources for each MASS and disseminates the policy to all MASSs and UAVs.

\emph{(4) Offloading Execution: }Based on the received policies, each MASS offloads its tasks to UAVs or the LEO satellite for further processing.

\section{Problem formulation}

\textcolor{b}{In this section, we formulate the research problem for minimizing the cumulative energy consumption of the double-edge-assisted computation offloading system, encompassing the energy expenditure of all the MASSs, UAVs, and the LEO satellite. The primary goal is to optimize the energy efficiency of the system while ensuring that each component operates within its respective constraints, particularly focusing on the energy usage of the UAVs, MASSs, and LEO satellite, as well as meeting the required latency constraints.}

\textcolor{b}{Based on the offloading model illustrated in Fig. 2, the system architecture consists of multiple MASSs, UAVs, and an LEO satellite working collaboratively to handle computation-intensive tasks. The MASSs, which are often resource-constrained in terms of computing resource and energy, offload their tasks to nearby UAVs or the LEO satellite for further processing. The UAVs act as intermediate edge computing nodes, providing additional computing resources closer to the MASSs, while the LEO satellite serves as a high-altitude computing platform with broader coverage and significant computational capabilities.} 

\textcolor{b}{Our objective is to minimize the system energy dissipation while adhering to latency requirement of each MASS, by jointly optimizing the offloading decision matrix $\bm{a}=\left\{a_{mn}\right\}_{m\in\mathcal{M},n\in\mathcal{N}}$, the offloading volume matrix $\bm{s}=\left\{s_{mn}\right\}_{m\in\mathcal{M},n\in\mathcal{N}}$, the computing resource allocation matrix of UAVs $\boldsymbol{\rho}^U=\left\{\rho_{mn}^U\right\}_{m\in\mathcal{M},n\in\mathcal{N}}$, and the computing resource allocation matrix of the LEO satellite $\boldsymbol{\rho}^L=\left\{\rho_{mn}^L\right\}_{m\in\mathcal{M},n\in\mathcal{N}}$, respectively. 
The system energy consumption minimization problem is formulated as}

\textbf{(P0):}
\begin{displaymath}
\min_{\bm{a},\bm{s},\boldsymbol{\rho}^U,\boldsymbol{\rho}^L}E^{tot} \tag{25}
\end{displaymath}

\begin{subequations}
\begin{align}
s.t.\nonumber\\
&0 \leq a_{mn}\leq 1, \forall m\in \mathcal{M},\forall n\in \mathcal{N}, \\
&0 \leq s_{mn}\leq S_{mn}, \forall m\in \mathcal{M},\forall n\in \mathcal{N},\\
&T_{mn}^{tot}\leq T_{mn}^{\max}, \forall m\in \mathcal{M},\forall n\in \mathcal{N},\\
&t_{mn}^L+T_{mn}^L\leq T^{\max}, \forall m\in \mathcal{M},\forall n\in \mathcal{N},\\
&\vert\vert\textbf{q}_m-\textbf{q}_{mn}\vert\vert\leq d^{\max}, \forall m\in \mathcal{M},\forall n\in \mathcal{N},\\
&\sum_{n\in\mathcal{N}}\rho_{mn}^U\leq \rho_m^{\max},\forall n\in \mathcal{N},\\
&\sum_{m\in\mathcal{M}}\sum_{n\in\mathcal{N}}\rho_{mn}^L\leq \rho_{\max}^L,\forall m\in \mathcal{M}, \forall n\in \mathcal{N},\\
&\rho_{mn}^l\geq0,\rho_{mn}^U\geq0,\rho_{mn}^L\geq0,\forall m\in \mathcal{M},\forall n\in \mathcal{N},\\
&p_{mn}^U\leq P_{\max}^U, \forall m\in \mathcal{M},\forall n\in \mathcal{N},\\
&p_{mn}^L\leq P_{\max}^L, \forall m\in \mathcal{M},\forall n\in \mathcal{N},\\
&\sum_{n\in\mathcal{N}}E_{mn}^U\leq E_m^{\max},\forall n\in \mathcal{N},\\
&\sum_{m\in\mathcal{M}}\sum_{n\in\mathcal{N}}E_{mn}^L\leq E_{\max}^L,\forall m\in \mathcal{M},\forall n\in \mathcal{N}.
\end{align} 
\end{subequations}
In \textbf{(P0)}, constraint (25a) denotes that the offloading ratio of $M_{mn}$ is between 0 and 1, constraint (25b) indicates that the uploading data of $M_{mn}$ cannot exceed the total workloads $S_{mn}$, constraint (25c) guarantees a delay bound of $M_{mn}$ for task offloading,
constraint (25d) provides the maximum latency guarantee for offloading task workloads to the LEO satellite,
constraint (25e) guarantees the maximum communication distance between $M_{mn}$ and $U_m$,
constraints (25f) and (25g) indicate the total computational capability of UAVs and the LEO satellite is bounded by $\rho_m^{\max}$ and $\rho_{\max}^L$, respectively,
constraint (25i) ensures that the transmission power to $U_m$ cannot exceed the maximum $P_{\max}^U$, constraint (25j) provides the maximum transmission power guarantee for offloading partial workloads to the LEO satellite, 
and constraints (25k) and (25l) indicate the total energy consumption of $U_m$ and the LEO satellite is bounded by $E_m^{\max}$ and $E_{\max}^L$, respectively.

\section{Energy-Efficient Double-Edge-Assisted Task Offloading Framework}
In \textbf{(P0)}, there are four sets of optimization variables, namely, the offloading mode decisions, the offloading volume decisions, the computing resource allocation decisions of UAVs, and the computing resource allocation decisions of the LEO satellite. 
%The optimization of the former two variables is intimately tied to the offloading mode, and the optimization of the last two variables revolves around the strategic computing resource allocation of UAVs and LEO satellite. 
To achieve an optimal solution with minimal computation overhead, the AO method\ccc{9978929} is employed and a layered approach is proposed to approximate the solution efficiently. We first optimize the offloading mode and the offloading volume decisions by fixing the computing resource allocation of UAVs and the LEO satellite with a multi-round iterative search algorithm. Then, we decompose the joint optimization problem of the computing resource allocation of UAVs and the LEO satellite by exploiting the convex structure and applying the Lagrangian dual decomposition method. The proposed solution framework is shown in Fig. 3. 
\begin{figure}[htbp]
\centerline{\includegraphics[width=0.45\textwidth]{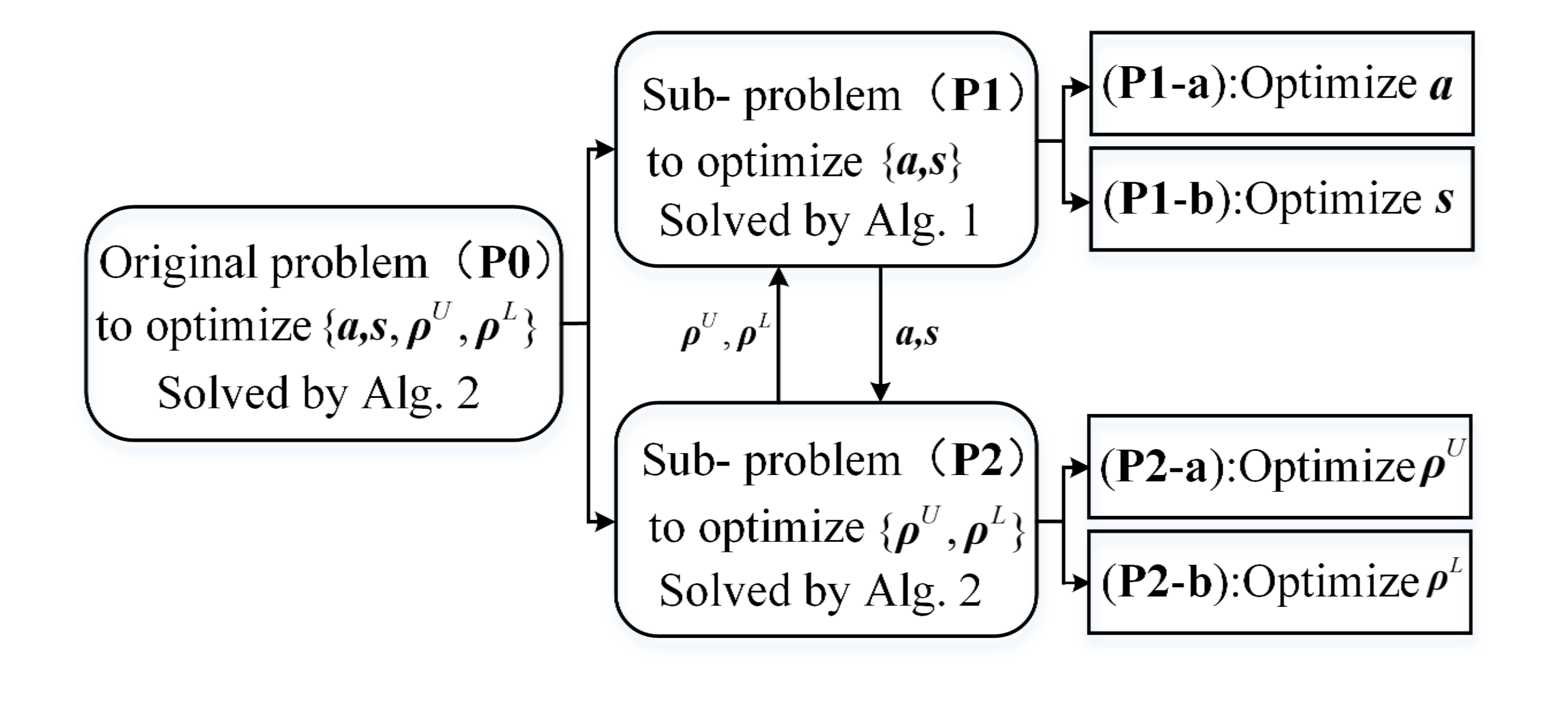}}
\caption{The proposed solution approach.}
\label{fig}
\end{figure}

\subsection{Joint Optimization of Offloading Mode and Volume}
Given the computing resource allocation of UAVs and the LEO satellite, i.e., fixing $\boldsymbol{\rho}^U$ and $\boldsymbol{\rho}^L$, \textbf{(P0)} is reformulated as

\textbf{(P1):}
\begin{displaymath}
\min_{\bm{a},\bm{s}}E^{tot}   \tag{26}
\end{displaymath}
\begin{subequations}
\begin{align}	
s.t.\,\ (25a)\sim (25d), (25i)\sim (25l).
\end{align}
\end{subequations}
With (25) and constraint (25c), we obtain the upper and lower bounds of $a_{mn}$ as
\begin{equation}
a_{mn}^{U}=\min\left\{1,\frac{\rho_{mn}^U\left(T_{mn}^{\max}-t_{mn}^U\right)}{s_{mn}c_m}\right\},
\end{equation}
and
\begin{equation}
a_{mn}^{L}=\max\left\{0,1-\frac{\rho_{mn}^L\left(T_{mn}^{\max}-t_{mn}^L\right)}{s_{mn}c^L}\right\}.
\end{equation}
Similarly, the upper and lower bounds of $s_{mn}$ are denoted, respectively, as
\begin{align}
&s_{mn}^{U}=\\ \nonumber
&\min\left\{S_{mn},\frac{\rho_{mn}^U\left(T_{mn}^{\max}-t_{mn}^U\right)}{a_{mn}c_m},\frac{\rho_{mn}^L\left(T_{mn}^{\max}-t_{mn}^L\right)}{\left(1-a_{mn}\right)c^L}\right\},
\end{align}
and
\begin{equation}
s_{mn}^{L}=\max\left\{0,S_{mn}-\frac{T_{mn}^{\max}\rho_{mn}^l}{c_{mn}}\right\}.
\end{equation}

\emph{(1) Offloading Mode Optimization: }

We first optimize the offloading mode decision matrix $\bm{a}$, while fixing the offloading volume decision matrix $\bm{s}$, the computing resource allocation matrix $\boldsymbol{\rho}^U$ of UAVs, and the computing resource allocation matrix $\boldsymbol{\rho}^L$ of the LEO satellite, yielding

\textbf{(P1-a):}
\begin{displaymath}
\min_{\bm{a}}E^{tot}  \tag{31}
\end{displaymath}
\begin{subequations}
\begin{align}
s.t.\,\ (25a), (25c), (25d), (25i)\sim (25l).
\end{align}
\end{subequations}
Let $E(a_{mn})=E^{tot}$, the first derivative of $E(a_{mn})$ with respect to $a_{mn}$ is expressed as
\begin{align}
&E^{'}(a_{mn})=\frac{\partial E(a_{mn})}{\partial a_{mn}}=\frac{\sigma^2s_{mn}\ln2}{g_{mn}^UW_{mn}^{U}}2^{\frac{a_{mn}s_{mn}}{t_{mn}^UW_{mn}^U}} \nonumber\\
&-\frac{t_{mn}^LN_0s_{mn}\ln2}{\mid h_{mn}^L\mid^2\left(t_{mn}^L-\frac{2d_{mn}^{L}}{c}\right)}2^{\frac{(1-a_{mn})s_{mn}}{\left(t_{mn}^L-\frac{2d_{mn}^{L}}{c}\right)W_{mn}^L}} \nonumber\\
 &+ P_{mn}^U\frac{s_{mn}c_m}{\rho_{mn}^U}- P_{mn}^L\frac{s_{mn}c^L}{\rho_{mn}^L}.
\end{align}
The second derivative of $E(a_{mn})$ with respect to $a_{mn}$ is expressed as
\begin{align}
&E^{''}(a_{mn})=\frac{\partial^2 E(a_{mn})}{\partial^2 a_{mn}}=\frac{\sigma^2}{t_{mn}^Ug_{mn}^U}\left(\frac{s_{mn}\ln2}{W_{mn}^{U}}\right)^22^{\frac{a_{mn}s_{mn}}{t_{mn}^UW_{mn}^U}} \nonumber\\
&+\frac{t_{mn}^LN_0}{\mid h_{mn}^L\mid^2W_{mn}^L}\left(\frac{-s_{mn}\ln2}{t_{mn}^L-\frac{2d_{mn}^{L}}{c}}\right)^22^{\frac{(1-a_{mn})s_{mn}}{\left(t_{mn}^L-\frac{2d_{mn}^{L}}{c}\right)W_{mn}^L}}.
\end{align}
As $E^{''}(a_{mn})\geq 0$, $E(a_{mn})$ is convex with respect to $a_{mn}$, and the first derivative $E^{'}(a_{mn})$ increases with $a_{mn}$ in the interval $\left[a_{mn}^{L},a_{mn}^{U}\right]$.
%Then, we obtain the minimum $E(a_{mn})$ as
%\begin{equation}
%\min\left\{E(a_{mn})\right\}=
%\begin{cases}
%E\left(a_{mn}^{L}\right),& \text{$E^{'}(a_{mn}^{L})>0$,} \\
%E\left(a_{mn}^*\right),& \text{$E^{'}(a_{mn}^{L})\leq 0\leq E^{'}(a_{mn}^{U})$,} \\
%E\left(a_{mn}^{U}\right),& \text{$ E^{'}(a_{mn}^{U})<0$,}
%\end{cases}
%\end{equation}
Then, we obtain the optimal offloading decision $a_{mn}^*$ as
\begin{equation}
	a_{mn}^*=
	\begin{cases}
		a_{mn}^{L},& \text{$E^{'}(a_{mn}^{L})>0$,} \\
		a_{mn}^{\Delta},& \text{$E^{'}(a_{mn}^{L})\leq 0\leq E^{'}(a_{mn}^{U})$,} \\
	    a_{mn}^{U},& \text{$ E^{'}(a_{mn}^{U})<0$,}
	\end{cases}
\end{equation}
where $a_{mn}^{\Delta}$ is the root of $E^{'}(a_{mn})=0$. Here, we propose a multi-round iterative search (MRIS) algorithm to obtain the value of $a_{mn}^*$, as shown in Algorithm 1.

\begin{algorithm}
	\SetAlgoLined
	\caption{MRIS Algorithm.}\label{algorithm}
	\KwIn{ Given the tolerable computation-error $\delta$;}
	\KwOut{The optimal value $\left\{a_{mn}^*\right\}$; }
	\textbf{Initialization:} Set the current best solutions of $\left\{a_{mn}^*\right\}=\varnothing$\;
	Calculate the upper bound $a_{mn}^{U}$ with Eq. (27)\;
	Calculate the lower bound $a_{mn}^{L}$ with Eq. (28)\;
	\If{$E^{'}(a_{mn}^{L})>0$}{Set $a_{mn}^*=a_{mn}^{L}$\;}
	\If{$E^{'}(a_{mn}^{U})<0$}{Set $a_{mn}^*=a_{mn}^{U}$\;}
	\If{$E^{'}(a_{mn}^{L})\leq 0\leq E^{'}(a_{mn}^{U})$}{
	\While{$\vert a_{mn}^{U}-a_{mn}^{L}\vert>\delta$}{
	Update the current value of $a_{mn}^{cur}=\frac{1}{2}(a_{mn}^{U}+a_{mn}^{L})$\;
	Calculate the value of $E^{'}(a_{mn}^{cur})$ with Eq. (32)\;
		\eIf{ $E^{'}(a_{mn}^{cur})<0$}
	{
		Update the lower bound of the search range as $a_{mn}^{L}=a_{mn}^{cur}$\;
	}{
		\eIf{$E^{'}(a_{mn}^{cur})>0$}
		{
			Update the upper bound of the search range as $a_{mn}^{U}=a_{mn}^{cur}$\;
		}
		{Set $a_{mn}^*=a_{mn}^{cur}$\;}
	  }
	 }
	}
	
\end{algorithm}

\emph{(2) Offloading Volume Optimization: }

We then optimize the offloading volume matrix $\bm{s}$, while fixing the offloading decision matrix $\bm{a}$, the computing resource allocation matrix $\boldsymbol{\rho}^U$ of UAVs, and the computing resource allocation matrix  $\boldsymbol{\rho}^L$ of LEO satellite, yielding

\textbf{(P1-b):}
\begin{displaymath}
\min_{\bm{s}}E^{tot}  \tag{35}
\end{displaymath}
\begin{subequations}
\begin{align}
s.t.\,\ (25b)\sim(25d), (25i)\sim (25l).
\end{align}
\end{subequations}
Let $E(s_{mn})=E^{tot}$, the first derivative of $E(s_{mn})$ with respect to $s_{mn}$ is expressed as
\begin{align}
&E^{'}(s_{mn})=\frac{\partial E(s_{mn})}{\partial s_{mn}}=-P_{mn}^l\frac{c_{mn}}{\rho_{mn}^l}+ \frac{\sigma^2a_{mn}\ln2}{g_{mn}^UW_{mn}^{U}}2^{\frac{a_{mn}s_{mn}}{t_{mn}^UW_{mn}^U}} \nonumber\\
&+\frac{t_{mn}^LN_0(1-a_{mn})\ln2}{\mid h_{mn}^L\mid^2\left(t_{mn}^L-\frac{2d_{mn}^{L}}{c}\right)}2^{\frac{(1-a_{mn})s_{mn}}{\left(t_{mn}^L-\frac{2d_{mn}^{L}}{c}\right)W_{mn}^L}} \nonumber\\
 &+ P_{mn}^U\frac{a_{mn}c_m}{\rho_{mn}^U}+ P_{mn}^L\frac{(1-a_{mn})c^L}{\rho_{mn}^L}.
\end{align}
The second derivative of $E(s_{mn})$ with respect to $s_{mn}$ is expressed as
\begin{align}
&E^{''}(s_{mn})=\frac{\partial^2 E(s_{mn})}{\partial^2 s_{mn}}=\frac{\sigma^2}{t_{mn}^Ug_{mn}^U}\left(\frac{a_{mn}\ln2}{W_{mn}^{U}}\right)^22^{\frac{a_{mn}s_{mn}}{t_{mn}^UW_{mn}^U}} \nonumber\\
&+\frac{t_{mn}^LN_0}{\mid h_{mn}^L\mid^2W_{mn}^L}\left(\frac{(1-a_{mn})\ln2}{t_{mn}^L-\frac{2d_{mn}^{L}}{c}}\right)^22^{\frac{(1-a_{mn})s_{mn}}{\left(t_{mn}^L-\frac{2d_{mn}^{L}}{c}\right)W_{mn}^L}}.
\end{align}
As $E^{''}(s_{mn})\geq 0$, $E(s_{mn})$ is convex with respect to $s_{mn}$, and the first derivative $E^{'}(s_{mn})$ is increasing with $s_{mn}$ in the interval $\left[s_{mn}^{L},s_{mn}^{U}\right]$.
%Then, we obtain the minimum $E(s_{mn})$ as
%\begin{equation}
%\min\left\{E(s_{mn})\right\}=
%\begin{cases}
%E\left(s_{mn}^{L}\right),& \text{$E^{'}(s_{mn}^{L})>0$,} \\
%E\left(s_{mn}^*\right),& \text{$E^{'}(s_{mn}^{L})\leq 0\leq E^{'}(s_{mn}^{U})$,} \\
%E\left(s_{mn}^{U}\right),& \text{$ E^{'}(s_{mn}^{U})<0$,}
%\end{cases}
%\end{equation}
Then, we obtain the optimal $s_{mn}^*$ as
\begin{equation}
	s_{mn}=
	\begin{cases}
		s_{mn}^{L},& \text{$E^{'}(s_{mn}^{L})>0$,} \\
		s_{mn}^{\Delta},& \text{$E^{'}(s_{mn}^{L})\leq 0\leq E^{'}(s_{mn}^{U})$,} \\
		s_{mn}^{U},& \text{$ E^{'}(s_{mn}^{U})<0$,}
	\end{cases}
\end{equation}
where $s_{mn}^{\Delta}$ is the root of $E^{'}(s_{mn})=0$. Similarly, we employ the MRIS algorithm to obtain the value of $s_{mn}^*$.

\subsection{Joint Optimization of Computing Resource Allocation}

Given the offloading decision matrix $\bm{a}$ and offloading volume matrix $\bm{s}$, \textbf{(P0)} is reformulated as

\textbf{(P2):}
\begin{displaymath}
\min_{\boldsymbol{\rho}^U,\boldsymbol{\rho}^L}E^{tot} \tag{39}
\end{displaymath}
\begin{subequations}
\begin{align}
s.t.\,\ (25c), (25d), (25f), (25g), (25k), (25l).
\end{align}
\end{subequations}
\textbf{(P2)} can be decomposed into the following two sub-problems. 

\emph{(1) Optimization of UAV Computing Resource Allocation: }

We first optimize the computing resource allocation matrix $\boldsymbol{\rho}^U$ of UAVs, given the offloading mode decision matrix $\bm{a}$, the offloading volume decision matrix $\bm{s}$, and the computing resource allocation matrix $\boldsymbol{\rho}^L$ of the LEO satellite, yielding

\textbf{(P2-a):}
\begin{displaymath}
\min_{\boldsymbol{\rho}^U}E^{tot}  \tag{40}
\end{displaymath}
\begin{subequations}
\begin{align}
s.t. \,\ &(25f),\nonumber\\
	&t_{mn}^U+\frac{a_{mn}s_{mn}c_m}{\rho_{mn}^U} \leq T_{mn}^{\max}, \forall m\in \mathcal{M},\forall n\in \mathcal{N}, \\
	&\sum_{n\in\mathcal{N}}P_{mn}^U\frac{a_{mn}s_{mn}c_m}{\rho_{mn}^U}\leq E_m^{\max},\forall n\in \mathcal{N}.
\end{align} 
\end{subequations}
%\begin{equation}
%	s.t.\,\ (22c),(22f),(22k).
%\end{equation}

The constraints of \textbf{(P2-a)} are convex with respect to $\rho_{mn}^U$. The second derivative of the objective function $E^{tot}$ with respect to $\rho_{mn}^U$ is formulated as
\begin{equation}
\frac{\partial^2 E^{tot}(\rho_{mn}^U)}{\partial^2 \rho_{mn}^{U}}=\frac{2P_{mn}^Ua_{mn}s_{mn}c_m}{\rho_{mn}^{U3}}\geq 0.
\end{equation}
The objective function $E^{tot}$ is convex with respect to $\rho_{mn}^U$. Then, \textbf{(P2-a)} is formulated as a convex optimization problem, which is solved with the Karush-Kuhn-Tucker (KKT) conditions\ccc{10107791},\ccc{10391072}.
% \bm{a},\bm{s},\boldsymbol{\rho}^U,\boldsymbol{\rho}^L

Specifically, the Lagrangian function of \textbf{(P2-a)} is formulated by
\begin{align}
&\mathcal{L}(\boldsymbol{\rho}^U,\boldsymbol{\lambda}_1,\boldsymbol{\lambda}_2,\boldsymbol{\lambda}_3)  \nonumber\\
&=\sum_{m\in\mathcal{M}}\sum_{n\in\mathcal{N}}P_{mn}^l\frac{(S_{mn}-s_{mn})c_{mn}}{\rho_{mn}^l}+P_{mn}^U\frac{a_{mn}s_{mn}c_m}{\rho_{mn}^{U}}\nonumber\\
&+\frac{t_{mn}^U\sigma^2}{g_{mn}^U}\left(2^{\frac{a_{mn}s_{mn}}{t_{mn}^UW_{mn}^U}}-1\right)+P_{mn}^L\frac{(1-a_{mn})s_{mn}c^L}{\rho_{mn}^{L}}\nonumber\\
&+\frac{t_{mn}^LW_{mn}^LN_0}{\mid h_{mn}^L\mid^2}\left[2^{\frac{(1-a_{mn})s_{mn}}{\left(t_{mn}^L-\frac{2d_{mn}^{L}}{c}\right)W_{mn}^L}}-1\right]  \nonumber\\
&-\sum_{m\in\mathcal{M}}\sum_{n\in\mathcal{N}}\lambda_{mn}^{1}\left(t_{mn}^U+\frac{a_{mn}s_{mn}c_m}{\rho_{mn}^U}- T_{mn}^{\max}\right)  \nonumber\\
&-\sum_{m\in\mathcal{M}}\lambda_m^{2}\left(\sum_{n\in\mathcal{N}}\rho_{mn}^U-\rho_m^{\max}\right)  \nonumber\\
&-\sum_{m\in\mathcal{M}}\lambda_m^{3}\left(\sum_{n\in\mathcal{N}}P_{mn}^U\frac{a_{mn}s_{mn}c_m}{\rho_{mn}^U}-E_m^{\max}\right)
\end{align}
where $\boldsymbol{\lambda}_1=\left\{\lambda_{mn}^{1}\right\}$, $\boldsymbol{\lambda}_2=\left\{\lambda_m^{2}\right\}$, $\boldsymbol{\lambda}_3=\left\{\lambda_m^{3}\right\}$ are the non-negative Lagrange multipliers. The optimal computing resource allocation $\rho_{mn}^{U*}$ at $U_m$ and the optimal Lagrange multipliers should satisfy the following KKT conditions for $\forall m\in \mathcal{M},\forall n\in \mathcal{N}$, given by
\begin{align}
&\frac{\partial \mathcal{L}}{\partial \rho_{mn}^U}=\frac{-P_{mn}^Ua_{mn}s_{mn}c_m}{\rho_{mn}^{U*2}}+\lambda_{mn}^{1*}\frac{a_{mn}s_{mn}c_m}{\rho_{mn}^{U*2}} \nonumber\\
&-\lambda_m^{2*}+\lambda_m^{3*}\frac{P_{mn}^Ua_{mn}s_{mn}c_m}{\rho_{mn}^{U*2}}=0,
\end{align}
\begin{equation}
\sum_{m\in\mathcal{M}}\sum_{n\in\mathcal{N}}\lambda_{mn}^{1*}\left(t_{mn}^U+\frac{a_{mn}s_{mn}c_m}{\rho_{mn}^{U*}}- T_{mn}^{\max}\right)=0,
\end{equation}
\begin{equation}
\sum_{m\in\mathcal{M}}\lambda_m^{2*}\left(\sum_{n\in\mathcal{N}}\rho_{mn}^{U*}-\rho_m^{\max}\right)=0,
\end{equation}
\begin{equation}
\sum_{m\in\mathcal{M}}\lambda_m^{3*}\left(\sum_{n\in\mathcal{N}}P_{mn}^U\frac{a_{mn}s_{mn}c_m}{\rho_{mn}^{U*}}-E_m^{\max}\right)=0.
\end{equation}
Based on Eq. (43)$\sim $Eq. (46), we obtain the value of $\rho_{mn}^{U*}$ as
\begin{equation}
\rho_{mn}^{U*}=\frac{a_{mn}s_{mn}c_m}{T_{mn}^{\max}-T_{mn}^{Ut}}.
\end{equation}

\emph{(2) Optimization of LEO Satellite Computing Resource Allocation: }

Then, we optimize the computing resource allocation matrix $\boldsymbol{\rho}^L$ of the LEO satellite, given the offloading mode decision matrix $\bm{a}$, the offloading volume decision matrix $\bm{s}$, and the computing resource allocation matrix $\boldsymbol{\rho}^U$ of UAVs, yielding

\textbf{(P2-b):}
\begin{displaymath}
\min_{\boldsymbol{\rho}^L}E^{tot} \tag{48}
\end{displaymath}
\begin{subequations}
\begin{align}
s.t. \,\ &(25g),\nonumber\\
	&t_{mn}^L+\frac{(1-a_{mn})s_{mn}c_m}{\rho_{mn}^L} \leq T_{mn}^{\max},\forall m\in \mathcal{M},\forall n\in \mathcal{N},  \\
	&t_{mn}^L+\frac{(1-a_{mn})s_{mn}c_m}{\rho_{mn}^L} \leq T^{\max},\forall m\in \mathcal{M},\forall n\in \mathcal{N},\\
	&\sum_{m\in\mathcal{M}}\sum_{n\in\mathcal{N}}P_{mn}^L\frac{a_{mn}s_{mn}c^L}{\rho_{mn}^L}\leq E_{\max}^L,\forall m\in \mathcal{M},\forall n\in \mathcal{N}.
\end{align} 
\end{subequations}
%\begin{equation}
%	s.t.\,\ (22c), (22d), (22g), (22l).
%\end{equation}
Similarly, the constraints of \textbf{(P2-b)} are convex with respect to $\rho_{mn}^L$. The second derivative of the objective function $E^{tot}$ with respect to $\rho_{mn}^L$ is formulated as
\begin{equation}
\frac{\partial^2 E^{tot}(\rho_{mn}^L)}{\partial^2 \rho_{mn}^{L}}=\frac{2P_{mn}^L(1-a_{mn})s_{mn}c^L}{\rho_{mn}^{L3}}\geq 0.
\end{equation}
We observe that $E^{tot}$ is also convex with respect to $\rho_{mn}^L$. Thus, \textbf{(P2-b)} is formulated as a convex optimization problem, which can be solved with KKT conditions.
% \bm{a},\bm{s},\boldsymbol{\rho}^U,\boldsymbol{\rho}^L
Specifically, we assume $T_{mn}^{\max}\leq T^{\max}$, the Lagrangian function of \textbf{(P2-b)} is expressed as
\begin{align}
&\mathcal{L}(\boldsymbol{\rho}^L,\boldsymbol{\mu}_1,\mu_2,\mu_3)  \nonumber\\
&=\sum_{m\in\mathcal{M}}\sum_{n\in\mathcal{N}}P_{mn}^l\frac{(S_{mn}-s_{mn})c_{mn}}{\rho_{mn}^l}+P_{mn}^U\frac{a_{mn}s_{mn}c_m}{\rho_{mn}^{U}}\nonumber\\
&+\frac{t_{mn}^U\sigma^2}{g_{mn}^U}\left(2^{\frac{a_{mn}s_{mn}}{t_{mn}^UW_{mn}^U}}-1\right)+P_{mn}^L\frac{(1-a_{mn})s_{mn}c^L}{\rho_{mn}^{L}}\nonumber\\
&+\frac{t_{mn}^LW_{mn}^LN_0}{\mid h_{mn}^L\mid^2}\left(2^{\frac{(1-a_{mn})s_{mn}}{\left(t_{mn}^L-\frac{2d_{mn}^{L}}{c}\right)W_{mn}^L}}-1\right)  \nonumber\\
&-\sum_{m\in\mathcal{M}}\sum_{n\in\mathcal{N}}\mu_{mn}^{1}\left(t_{mn}^L+\frac{(1-a_{mn})s_{mn}c^L}{\rho_{mn}^L}- T_{mn}^{\max}\right)  \nonumber\\
&-\mu_2\left(\sum_{m\in\mathcal{M}}\sum_{n\in\mathcal{N}}\rho_{mn}^L-\rho_{\max}^L\right)  \nonumber\\
&-\mu_3\left(\sum_{m\in\mathcal{M}}\sum_{n\in\mathcal{N}}P_{mn}^L\frac{(1-a_{mn})s_{mn}c^L}{\rho_{mn}^L}-E_{\max}^L\right)
\end{align}
where $\boldsymbol{\mu}_1=\left\{\mu_{mn}^{1}\right\}$, $\mu_2$ and $\mu_3$ are the non-negative Lagrange multipliers. The optimal computing resource allocation $\rho_{mn}^{L*}$ of the LEO satellite and the optimal Lagrange multipliers should satisfy the following KKT conditions for $\forall m\in \mathcal{M},\forall n\in \mathcal{N}$, given by
\begin{align}
&\frac{\partial \mathcal{L}}{\partial \rho_{mn}^L}=\frac{-P_{mn}^L(1-a_{mn})s_{mn}c^L}{\rho_{mn}^{L*2}}+\mu_{mn}^{1*}\frac{(1-a_{mn})s_{mn}c^L}{\rho_{mn}^{L*2}} \nonumber\\
&-\mu_2^{*}+\mu_3^{*}\frac{P_{mn}^L(1-a_{mn})s_{mn}c^L}{\rho_{mn}^{L*2}}=0,
\end{align}
\begin{equation}
\sum_{m\in\mathcal{M}}\sum_{n\in\mathcal{N}}\mu_{mn}^{1*}\left(t_{mn}^L+\frac{(1-a_{mn})s_{mn}c^L}{\rho_{mn}^{L*}}- T_{mn}^{\max}\right)=0,
\end{equation}
\begin{equation}
\mu_2^{*}\left(\sum_{m\in\mathcal{M}}\sum_{n\in\mathcal{N}}\rho_{mn}^{L*}-\rho_{\max}^L\right)=0,
\end{equation}
\begin{equation}
\mu_3^{*}\left(\sum_{m\in\mathcal{M}}\sum_{n\in\mathcal{N}}P_{mn}^L\frac{(1-a_{mn})s_{mn}c^L}{\rho_{mn}^{L*}}-E_{\max}^L\right)=0.
\end{equation}
Then, we obtain the value of $\rho_{mn}^{L*}$ as
\begin{equation}
\rho_{mn}^{L*}=\frac{(1-a_{mn})s_{mn}c^L}{T_{mn}^{\max}-T_{mn}^{Lt}}.
\end{equation}

%Based on the above derivations, a \underline{J}oint \underline{O}ptimization of the \underline{C}omputation \underline{R}esource \underline{A}llocation (JOCRA) of UAVs and the LEO satellite is elaborated in Algorithm 2.  

Based on the above derivations, the solution to \textbf{(P0)} (STP) is articulated in Algorithm 2. 
\begin{algorithm}
	\SetAlgoLined
	\caption{STP Algorithm.}\label{algorithm}
	\textbf{Initialization:} Set the maximum number of iterations $T$, set the initial value as $t=0$ of the iterations\;
	\While{$t<T$}{
	Given $\boldsymbol{\rho}^U$ and $\boldsymbol{\rho}^L$, calculate $\bm{a}$ and $\bm{s}$ with Algorithm 1\;
	Given $\bm{a}$, $\bm{s}$ and $\boldsymbol{\rho}^L$, calculate $\boldsymbol{\rho}^U$ with Eq. (47)\;
	Given $\bm{a}$, $\bm{s}$ and $\boldsymbol{\rho}^U$, calculate $\boldsymbol{\rho}^L$ with Eq. (55)\;
	Update $t\leftarrow t+1$\;
	}
	Set $\bm{a}^*=\bm{a}$, $\bm{s}^*=\bm{s}$, $\boldsymbol{\rho}^{U*}=\boldsymbol{\rho}^U$, $\boldsymbol{\rho}^{L*}=\boldsymbol{\rho}^L$ \;
	\KwOut{The optimal value $\bm{a}^*$, $\bm{s}^*$, $\boldsymbol{\rho}^{U*}$ and $\boldsymbol{\rho}^{L*}$\; }
\end{algorithm}

\subsection{Complexity Analysis}
\textcolor{b}{As presented in Fig. 3, to solve \textbf{(P0)}, we propose a layered structure and decompose the original problem \textbf{(P0)} into two subproblems, \textbf{(P1)} and \textbf{(P2)}. \textbf{(P1)} jointly optimizes the offloading mode and volume, respectively, and \textbf{(P2)} jointly optimizes the computing resource allocation of the UAVs and the LEO satellite, respectively. Specifically, to solve \textbf{(P1)}, Algorithm 1 is proposed to find the optimal $\left\{a_{mn}\right\}_{m\in\mathcal{M},n\in\mathcal{N}}^*$ and  
$\left\{s_{mn}\right\}_{m\in\mathcal{M},n\in\mathcal{N}}^*$ for each MASS. 
We denote the number of iterations of Algorithm 1 as $K$. Then, we obtain the computation complexity of Algorithm 1 as $\mathcal{O}\left( N\log_2K\right)$ for $N$ MASSs. 
For Algorithm 2, the complexity of the computing resource allocation is $\mathcal{O}(N)$ for all MASSs.
Assuming that $T$ represents the number of iterations required for the algorithm to converge, the total computational complexity of the proposed Algorithm 2 is $\mathcal{O}\left( TN\log_2K\right)$.
Therefore, the proposed algorithms have low complexity, which shows good scalability.}

\section{Performance Evaluation}
In this section, we conduct numerical analysis to validate the effectiveness of the proposed algorithms. Specifically, we evaluate the impact of key parameters on the energy consumption and compare the proposed scheme with the following three benchmark schemes.
\begin{itemize}
%\item[$(a)$] 
%Pure local computing (PLC): In this scheme, All the MASSs process their tasks locally without task offloading, i.e., $s_{mn}=0$, $\forall m\in \mathcal{M},\forall n\in \mathcal{N}$. 
\item[$(a)$] 
Paired offloading of multiple tasks (POMT) scheme\ccc{8735850}: In this scheme, each MASS can only execute its task locally or entirely offload it to one edge server for processing. 
\item[$(b)$] 
Equal offloading scheme (EOS): Similar to the Round Robin method in\ccc{10024305} and\ccc{10439163}, in this scheme, $M_{mn}$, $U_m$, and the LEO satellite each complete an identical amount of workloads for every MASS.
\item[$(c)$] 
Even allocation of computing resource (EACR) scheme: In this scheme, the computing resources of $U_m$ and the LEO satellite are evenly allocated among all MASSs. 
\end{itemize}

 \subsection{System setup}
We conduct all the numerical analysis with MATLAB on a PC configured using a Core i7-10510U 1.80 GHz CPU and 8 GB of RAM. 
We consider a double-edge-assisted SAMIN comprised of one LEO satellite and four UAVs hovering in the air with the positions of $(125,125,100)$m, $(125,375,100)$m, $(375,125,100)$m, $(375,375,100)$m, respectively. Each UAV covers 5 MASSs which navigate autonomously. 
The LEO satellite is responsible for determining offloading strategies and computing resource allocation policies for all MASSs. 
We assume each $M_{mn}$ has a total task volume of 10 Mbits. Each MASS communicates with UAV via C-band, utilizing a channel bandwidth of 12 MHz, and each MASS communicates with the LEO satellite through Ka-band, employing a channel bandwidth of 15 MHz. The main parameters  are shown in Table I.

\begin{table}[htbp]
\centering
\caption{Simulation Parameter Settings}
\begin{tabular}{l|l}%l=left, r=right,c=center分别代表左对齐，右对齐和居中，字母的个数代表列数
\hline
\hline
Parameters &Values  \\ 
\hline
Maximum latency for processing tasks of $M_{mn}$  \\
($T_{mn}^{\max}$)  &$1$s  \\ 
\hline
Spectral power of the additive white Gaussian  \\
noise ($\sigma^2$) &$7.9e-9$ mW  \\
\hline
Transmission bandwidth of $M_{mn}$ to $U_m$ ($W_{mn}^U$)  &$12$ MHz  \\ 
\hline
Transmission time between $M_{mn}$ and $U_m$ ($t_{mn}^U$)  &$0.4$s  \\
\hline
Channel power gain exponent ($\chi$)  &$1$  \\
\hline
Elevation angle ($\theta$)  &$30^{\circ}$  \\
\hline
Path loss exponent ($\zeta$)  &$1.6$   \\
\hline
Transmission time between $M_{mn}$ and LEO \\
satellite ($t_{mn}^L$)  &$0.7$s  \\
\hline
Transmission bandwidth of $M_{mn}$ to LEO \\
satellite ($W_{mn}^L$)  &$15$ MHz  \\
\hline
Height of the LEO satellite ($h$)  &$784$ km  \\
\hline
Path loss exponent ($\gamma$)  &$2$ \\
\hline
Speed of light ($c$)  &$3\times10^8$m/s \\
\hline
Number of CPU cycles for processing one bit of \\
data by $M_{mn}$ ($c_{mn}$) &$1\times10^3$\\ 
\hline
Number of CPU cycles for processing one bit of \\
data by $U_m$ ($c_m$) &$1\times10^3$ \\ 
\hline
Number of CPU cycles for processing one bit of \\
data by the LEO satellite ($c^L$) &$1\times10^3$ \\ 
\hline
CPU computing capability of $M_{mn}$ ($\rho_{mn}^l$)  &$7\times10^9$ cycles/s \\
\hline
\hline
\end{tabular}
\end{table}

 \subsection{Performance evaluation and analysis}
Fig. 4 illustrates the total energy consumption and the overall latency associated with completing $M_{mn}$'s workloads under different values of $a_{mn}$ and iteration index with fixed $t_{mn}^U=0.4s$ and $t_{mn}^L=0.7s$, respectively. We observe that  both $E_{mn}^{tot}$ and $T_{mn}^{tot}$ converge.
Furthermore, as the value of $a_{mn}$ increases, the overall latency $T_{mn}^{tot}$ decreases, while the total energy consumption $E_{mn}^{tot}$ increases. This arises because an increase in $a_{mn}$ leads to an increased workloads for $U_m$, which in turn requires more energy to handle additional tasks. Conversely, the expedited transmission between $M_{mn}$ and $U_m$ helps reduce the overall latency.

\begin{figure}[htbp]\color{b}
	\begin{minipage}{0.49\linewidth}
		\vspace{3pt}
		\centerline{\includegraphics[width=\textwidth]{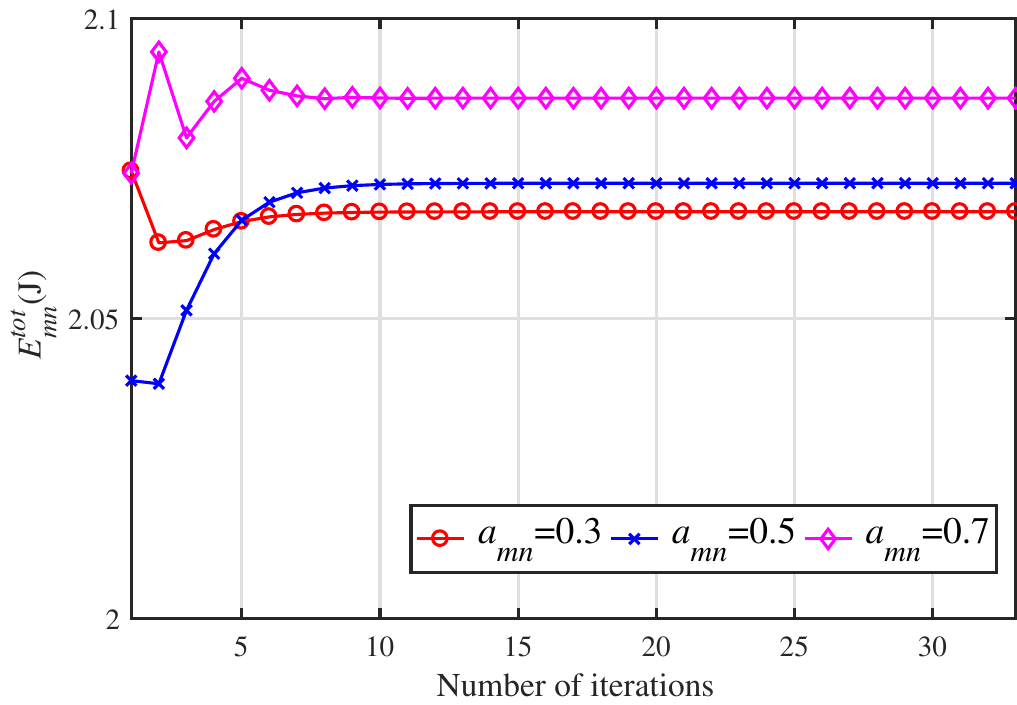}}
	 
		\centerline{(a) $E_{mn}^{tot}$ vs. $a_{mn}$}
	\end{minipage}
	\begin{minipage}{0.49\linewidth}
		\vspace{3pt}
		\centerline{\includegraphics[width=\textwidth]{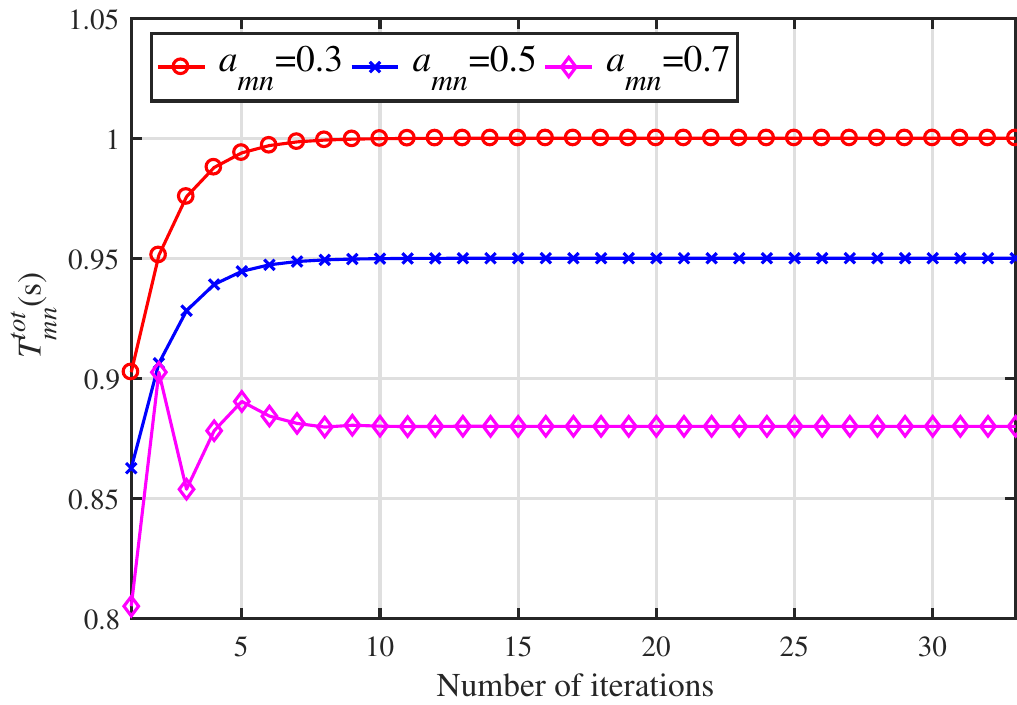}}
	 
		\centerline{(b) $T_{mn}^{tot}$ vs. $a_{mn}$}
	\end{minipage}
 
	\caption{The total energy consumption and overall latency associated with completing $M_{mn}$'s workloads under different values of $a_{mn}$ with fixed $t_{mn}^U=0.4s$ and $t_{mn}^L=0.7s$.}
	\label{fig}
\end{figure}

Fig. 5 illustrates the impact of transmission time (i.e., $t_{mn}^U$, $t_{mn}^L$)  on the total energy consumption $E_{mn}^{tot}$ and the overall latency $T_{mn}^{tot}$, respectively. 
We observe that both $E_{mn}^{tot}$ and $T_{mn}^{tot}$ converge as the number of iterations increases.
Specifically, Fig. 5(a) and Fig. 5(c) show that the total energy consumption decreases with increased transmission time, while Fig. 5(b) and Fig. 5(d) show that the overall latency increases with increased transmission time. This demonstrates the trade-off between energy consumption and latency. Within practical limits, a controlled increase in delay can lead to significant energy savings, ultimately enhancing the overall network performance.
 
\begin{figure*}[ht!]\color{b}
	
	\begin{minipage}{0.24\linewidth}%可修改0.32为其他比例，调整大小
		\vspace{1pt}
		%这个图片路径替换成你的图片路径即可使用
		\centerline{\includegraphics[width=\textwidth]{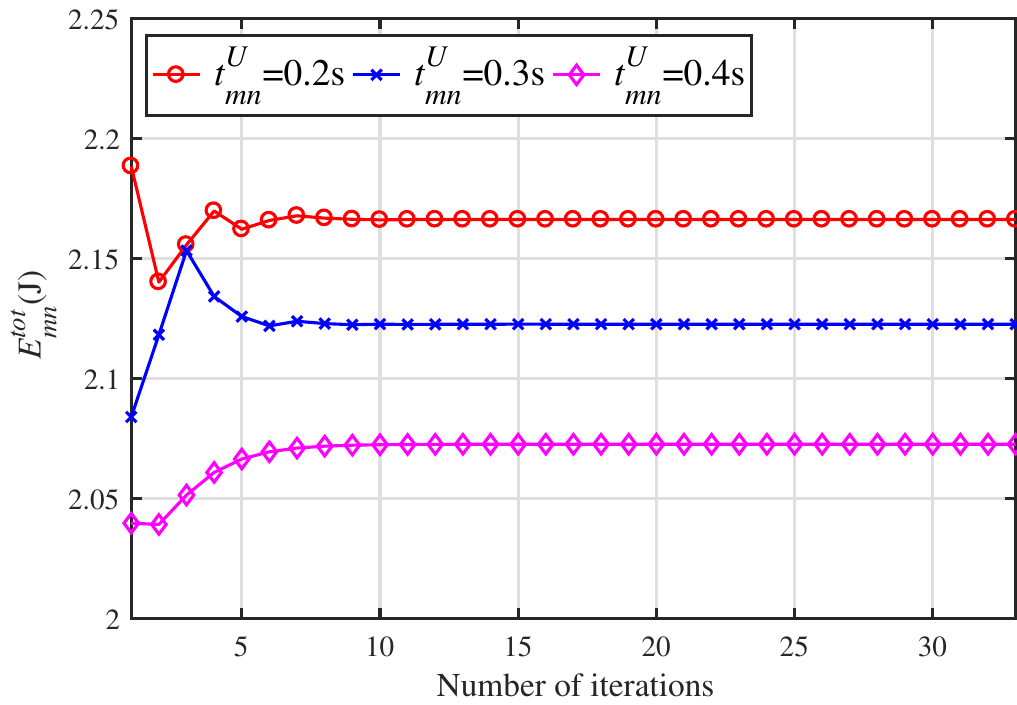}}
		% 加入对这列的图片说明
		\centerline{(a) $E_{mn}^{tot}$ vs. $t_{mn}^U$ }
	\end{minipage}
	\begin{minipage}{0.24\linewidth}
		\vspace{1pt}
		\centerline{\includegraphics[width=\textwidth]{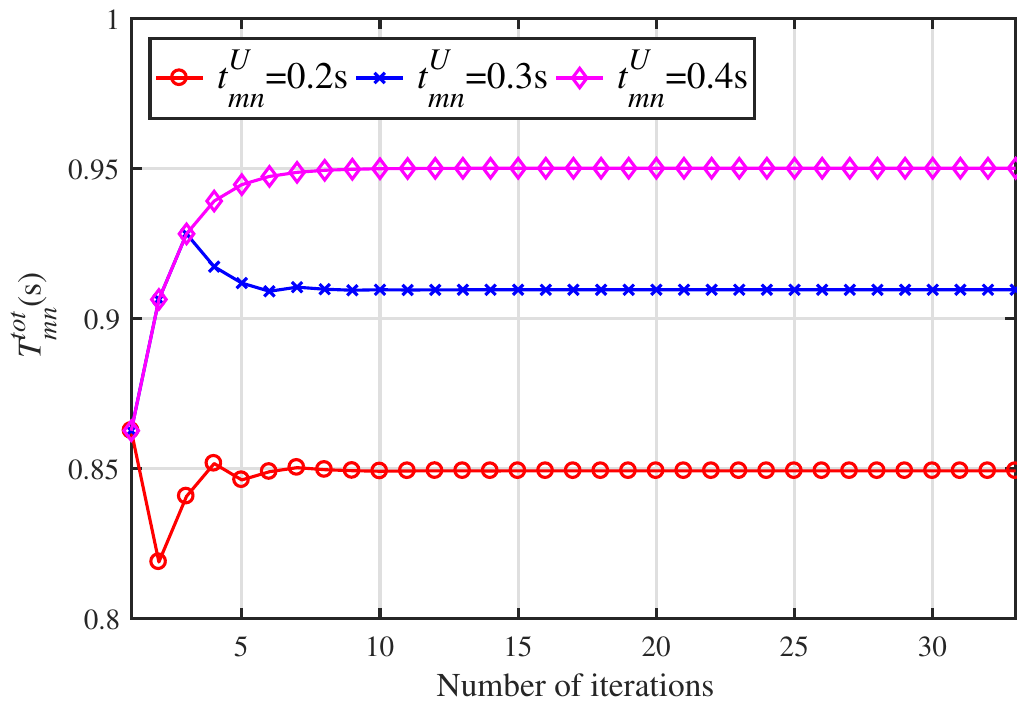}}
		
		\centerline{(b) $T_{mn}^{tot}$ vs. $t_{mn}^U$}
	\end{minipage}
	\begin{minipage}{0.24\linewidth}
		\vspace{1pt}
		\centerline{\includegraphics[width=\textwidth]{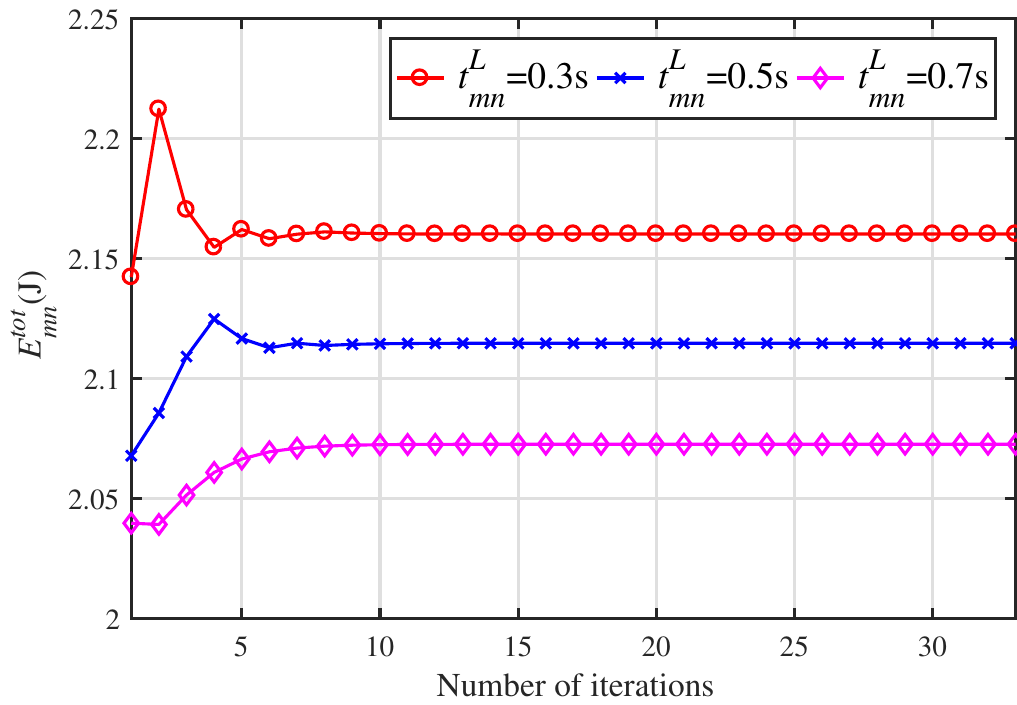}}
		
		\centerline{(c) $E_{mn}^{tot}$ vs. $t_{mn}^L$}
	\end{minipage}
	\begin{minipage}{0.24\linewidth}
		\vspace{1pt}
		\centerline{\includegraphics[width=\textwidth]{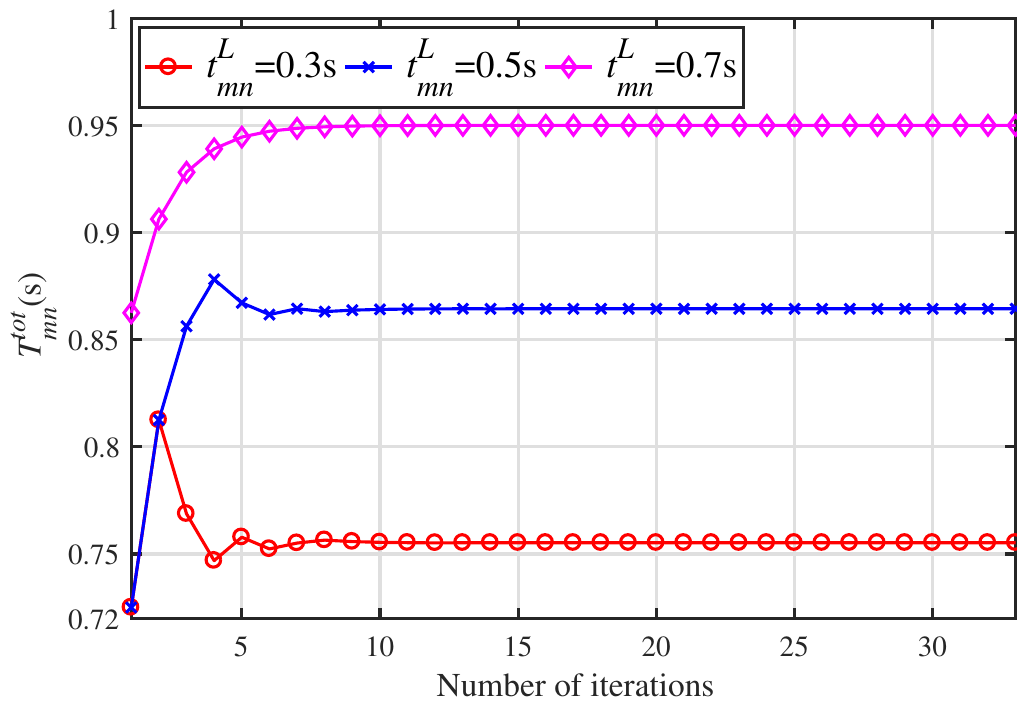}}
		
		\centerline{(d) $T_{mn}^{tot}$ vs. $t_{mn}^L$}
	\end{minipage}
	
	\caption{Illustration of total energy consumption and overall latency
		under different values of $t_{mn}^U$ and $t_{mn}^L$, respectively.}
	\label{fig}
\end{figure*} 
 
Fig. 6 illustrates the impact of the total task volume $S_{mn}$ on the offloading ratio $a_{mn}$, the total energy consumption $E_{mn}^{tot}$, and the overall latency $T_{mn}^{tot}$, respectively. Specifically, Fig. 6(a) highlights a decrease in $a_{mn}$ as $S_{mn}$ increases, due to the limitation in UAVs' computing capabilities, 
which requires the support of the LEO satellite to handle the additional computation workloads. Fig. 6(b) and Fig. 6(c) indicate that both $E_{mn}^{tot}$ and $T_{mn}^{tot}$  increase in tandem with the increase of $S_{mn}$. As $S_{mn}$ becomes higher, the MASSs, UAVs, and LEO satellite all require additional resources to execute the increased workloads, leading to a surge in energy expenditure and an extension of the overall latency.

\begin{figure*}[ht!]\color{b}
	
	\begin{minipage}{0.32\linewidth}%可修改0.32为其他比例，调整大小
		\vspace{2pt}
		%这个图片路径替换成你的图片路径即可使用
		\centerline{\includegraphics[width=\textwidth]{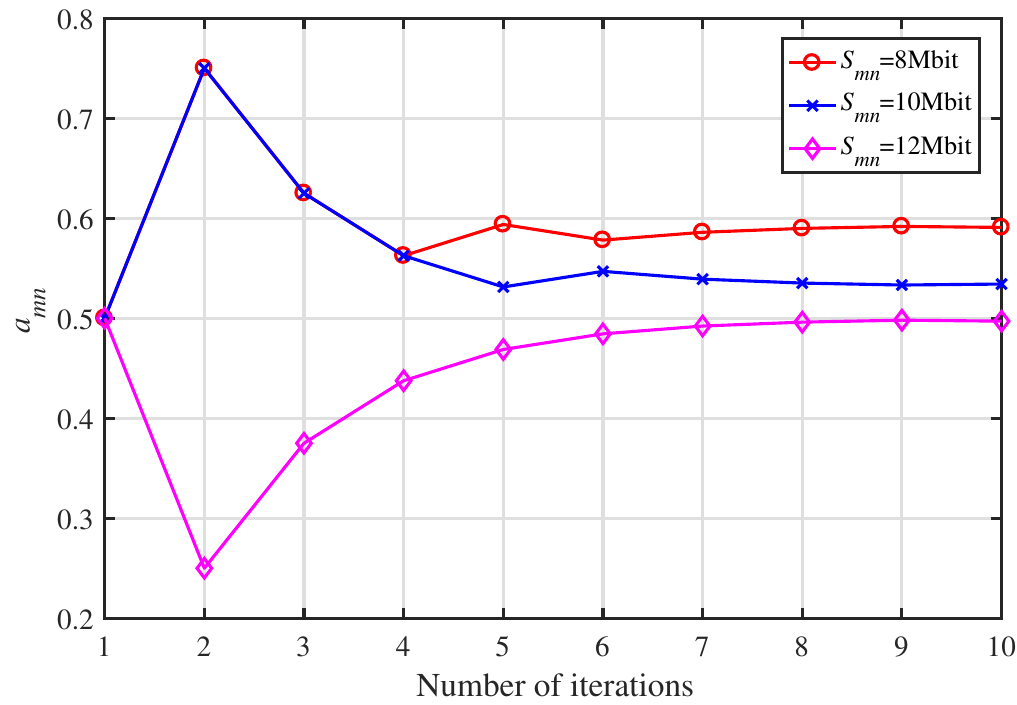}}
		% 加入对这列的图片说明
		\centerline{(a) $a_{mn}$ vs. $S_{mn}$ }
	\end{minipage}
	\begin{minipage}{0.32\linewidth}
		\vspace{2pt}
		\centerline{\includegraphics[width=\textwidth]{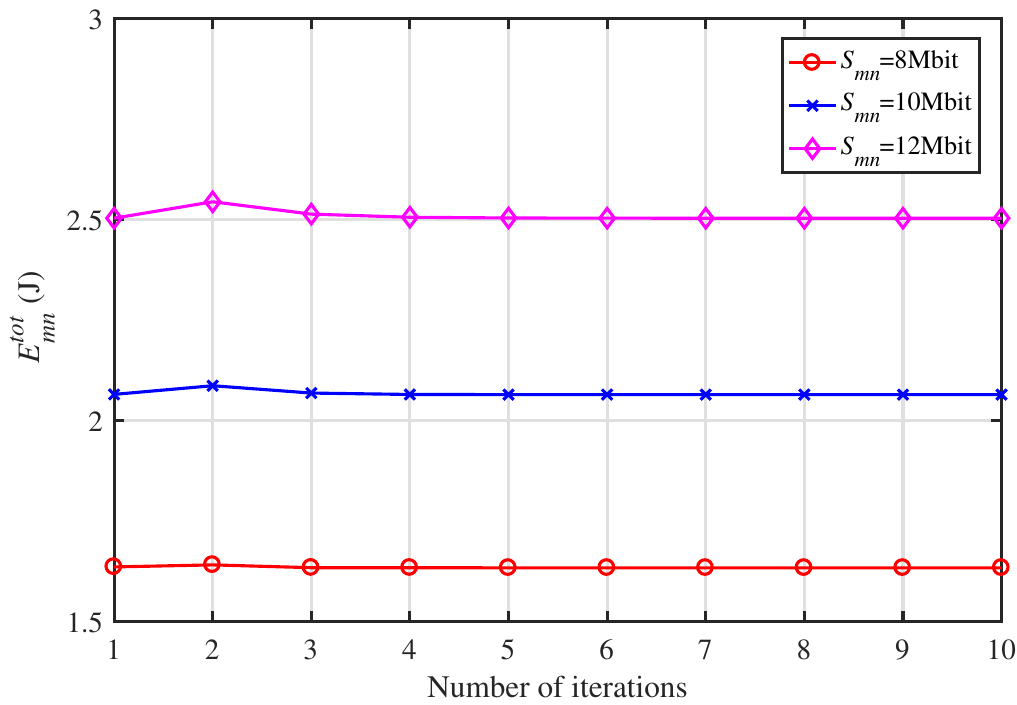}}
		
		\centerline{(b) $E_{mn}^{tot}$ vs. $S_{mn}$}
	\end{minipage}
	\begin{minipage}{0.32\linewidth}
		\vspace{2pt}
		\centerline{\includegraphics[width=\textwidth]{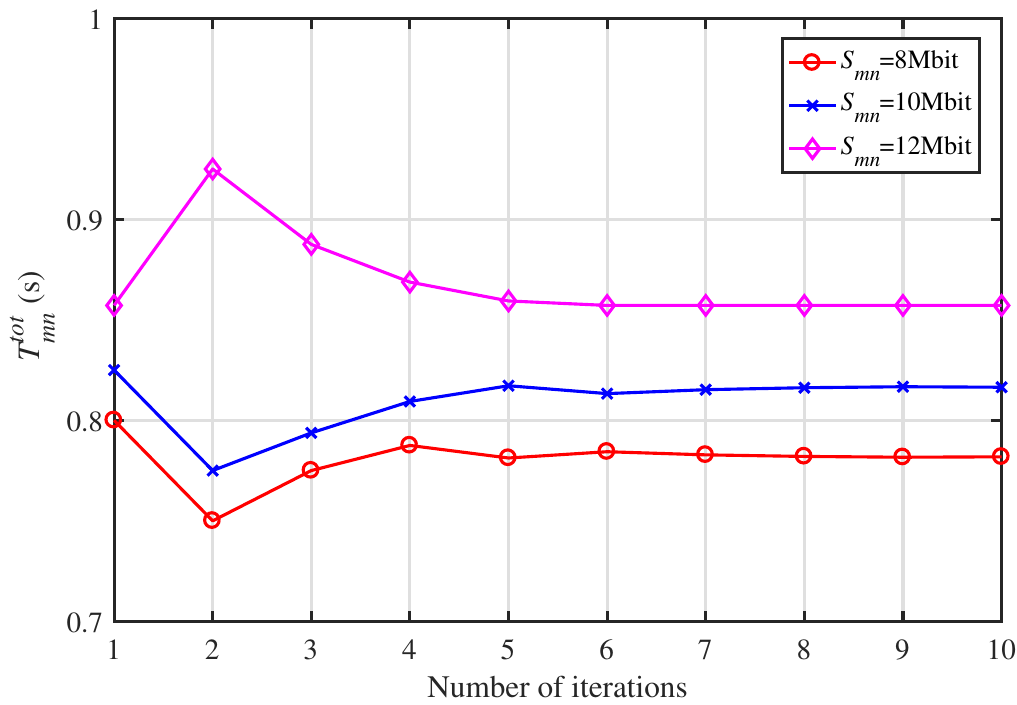}}
		
		\centerline{(c) $T_{mn}^{tot}$ vs. $S_{mn}$}
	\end{minipage}
	
	\caption{Illustration of offloading ratio, energy consumption and overall latency
		under different values of $S_{mn}$ and iteration index.}
	\label{fig}
\end{figure*}

\begin{figure*}[ht!]\color{b}
	
	\begin{minipage}{0.32\linewidth}%可修改0.32为其他比例，调整大小
		\vspace{2pt}
		%这个图片路径替换成你的图片路径即可使用
		\centerline{\includegraphics[width=\textwidth]{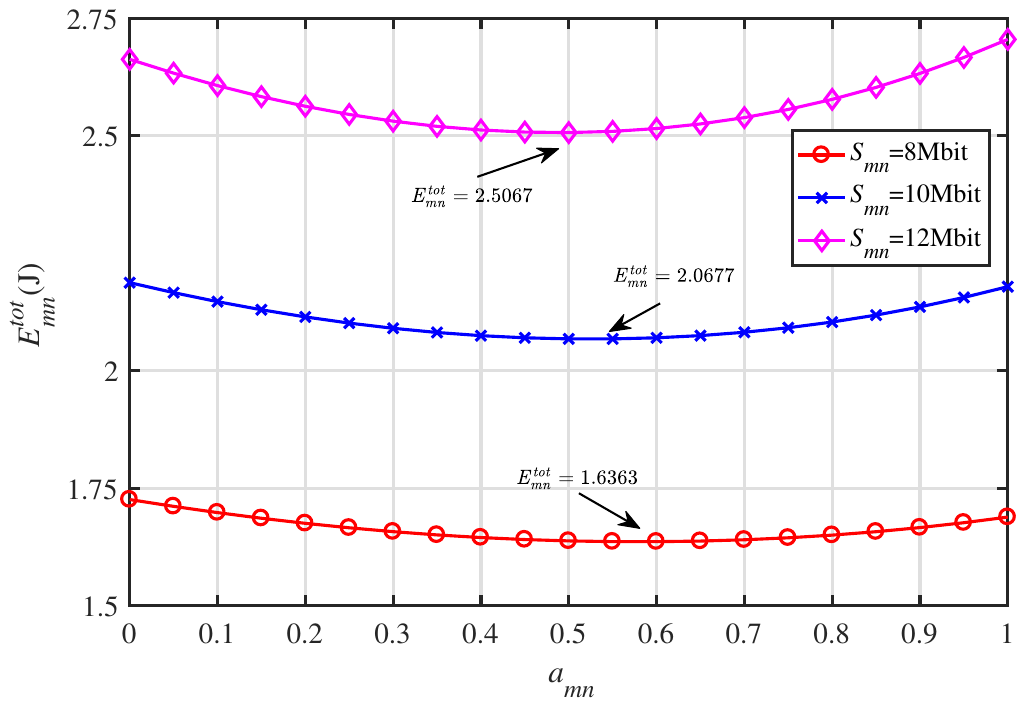}}
		% 加入对这列的图片说明
		\centerline{(a) $E_{mn}^{tot}$ vs. $a_{mn}$}
	\end{minipage}
	\begin{minipage}{0.32\linewidth}
		\vspace{2pt}
		\centerline{\includegraphics[width=\textwidth]{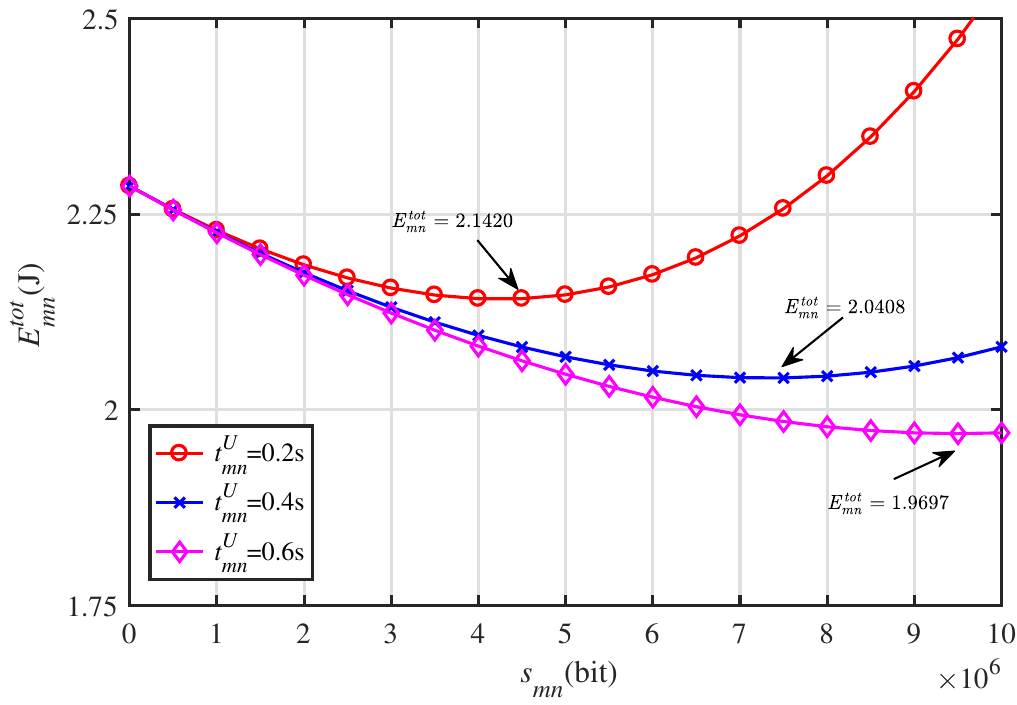}}
		
		\centerline{(b) $E_{mn}^{tot}$ vs. $s_{mn}$}
	\end{minipage}
	\begin{minipage}{0.32\linewidth}
		\vspace{2pt}
		\centerline{\includegraphics[width=\textwidth]{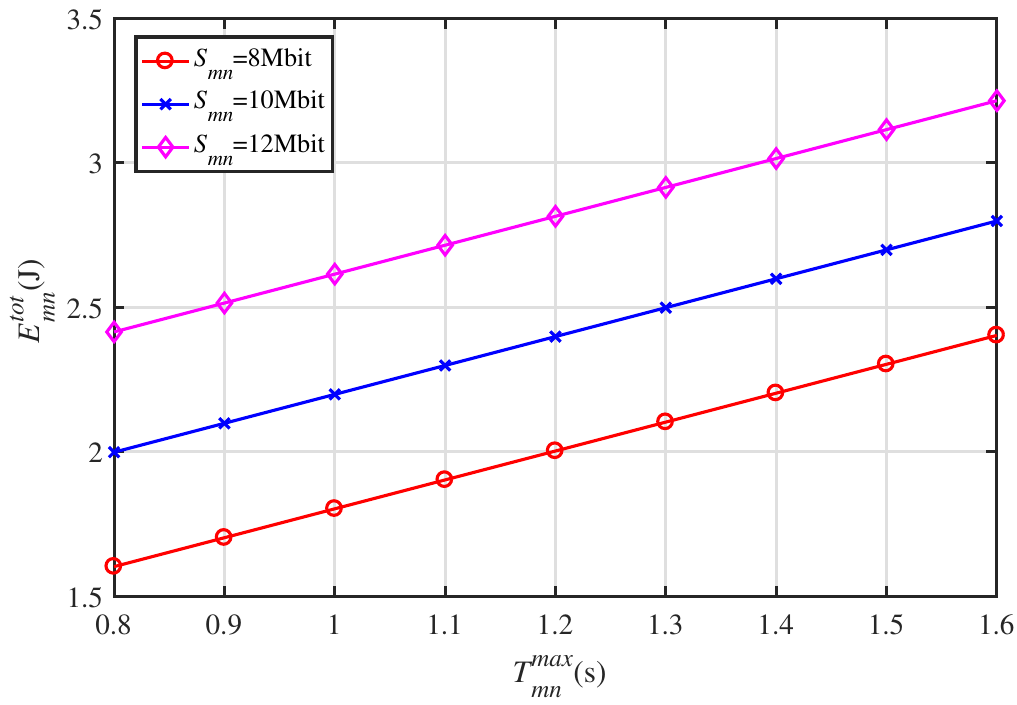}}
		
		\centerline{(c) $E_{mn}^{tot}$ vs. $T_{mn}^{\max}$}
	\end{minipage}
	
	\caption{The overall energy consumption under different values of $a_{mn}$, $s_{mn}$ and $T_{mn}^{\max}$.}
	\label{fig}
\end{figure*}

\begin{figure*}[ht!]\color{b}
	
	\begin{minipage}{0.32\linewidth}%可修改0.32为其他比例，调整大小
		\vspace{2pt}
		%这个图片路径替换成你的图片路径即可使用
		\centerline{\includegraphics[width=\textwidth]{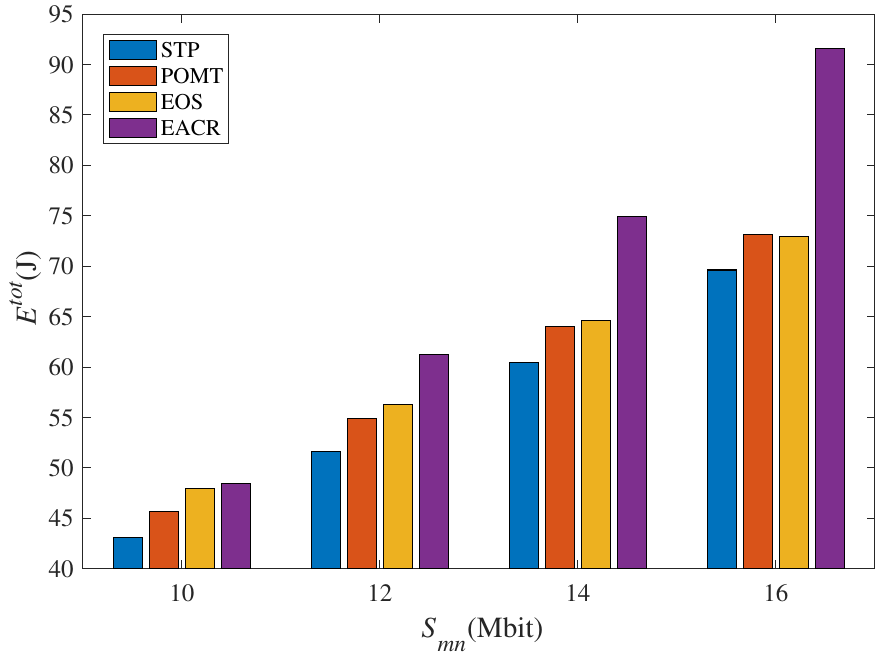}}
		% 加入对这列的图片说明
		\centerline{(a) $E^{tot}$ vs. $S_{mn}$ }
	\end{minipage}
	\begin{minipage}{0.32\linewidth}
		\vspace{2pt}
		\centerline{\includegraphics[width=\textwidth]{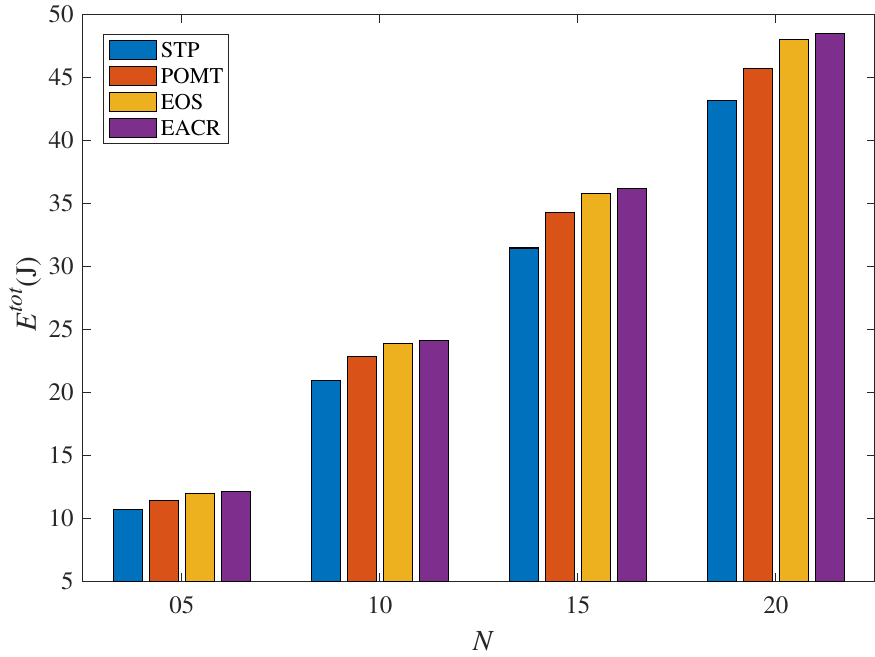}}
		
		\centerline{(b) $E^{tot}$ vs. $N$}
	\end{minipage}
	\begin{minipage}{0.32\linewidth}
		\vspace{2pt}
		\centerline{\includegraphics[width=\textwidth]{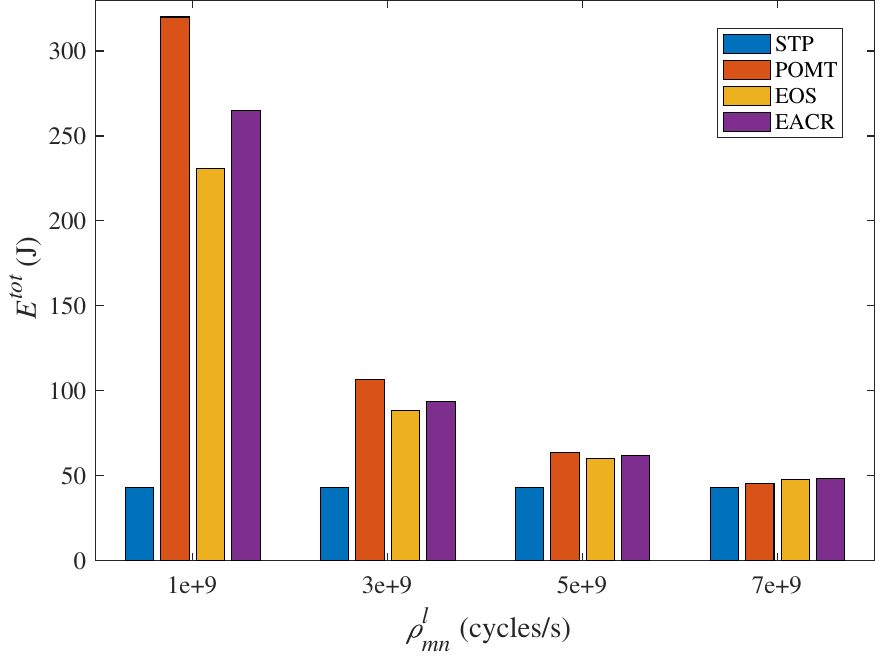}}
		
		\centerline{(c) $E^{tot}$ vs. $\rho_{mn}^l$}
	\end{minipage}
	
	\caption{Performance comparison of our proposed scheme with other benchmarks under different parameters (i.e., $S_{mn}$, $N$, $\rho_{mn}^l$).}
	\label{fig}
\end{figure*}

Fig. 7 demonstrates the overall energy consumption under different values of offloading ratio $a_{mn}$, offloading volume $s_{mn}$ and tolerable delay of $M_{mn}$, respectively. 
In Fig. 7(a), we observe that as $a_{mn}$ increases, the total energy consumption initially decreases, followed by a subsequent increase. The variation indicates the existence of an optimal value of $a_{mn}$, where the total energy consumption reaches its minimum. The reason for this stems from the workloads distribution between UAVs and the LEO satellite. When $a_{mn}$ is low, the LEO satellite undertakes a heavy workloads, resulting in increased total energy consumption. Conversely, as $a_{mn}$ increases, the workloads shifts to UAVs, which are more energy-efficient for specific tasks, thereby reducing the overall energy expenditure. It is essential to select an appropriate value of $a_{mn}$ to minimize total energy consumption.
Fig. 7(b) indicates the trend of $E_{mn}^{tot}$ with respect to the variation of $s_{mn}$ under different values of $t_{mn}^U$. 
As $s_{mn}$ increases, $E_{mn}^{tot}$ initially decreases and then rises, and there is always an optimal value of $s_{mn}$ that minimizes $E_{mn}^{tot}$. This is because when $s_{mn}$ is small, the workloads carried by $M_{mn}$ is substantial, leading to higher energy consumption. As $s_{mn}$ increases, the workloads carried by UAVs and the LEO satellite gradually increases, which helps reduce the total energy consumption. However, the value of $s_{mn}$ cannot increase indefinitely, as a larger $s_{mn}$ would lead to higher transmission energy consumption. 
Fig. 7(c) shows the impact of $M_{mn}$'s tolerable delay $T_{mn}^{\max}$ on the overall energy consumption. We see that $E_{mn}^{tot}$ undergoes an increase as $T_{mn}^{\max}$ escalates. In our formulated problem, $T_{mn}^{\max}$ represents a limiting factor. 
When the value of $T_{mn}^{\max}$ increases, both UAVs and the LEO satellite can fulfill the computation workloads within the prescribed time with reduced computation resources, which results in an increase in energy consumption according to \textcolor{b}{Eq. (17) and Eq. (21)}.

Fig. 8 illustrates the performance comparison of the proposed scheme with other three benchmarks under different parameters (i.e., the total workloads, the number of MASSs, and the computational capability of $M_{mn}$). Figs. 8(a), 8(b), and 8(c) clearly demonstrate the superiority of our proposed scheme, which effectively diminishes the overall energy consumption. The reason is that we determine the optimal computation offloading decisions and resource allocation strategies by minimizing the total energy consumption.

\section{Conclusion and Future Work}
In this paper, we have considered an SAMIN and proposed a double-edge assisted computation offloading scheme for MASSs by jointly optimizing the offloading mode, the offloading volume, the computing resource allocation of UAVs and the LEO satellite, respectively, to improve the efficiency of computation offloading. 
We define a scenario where both UAVs and the LEO satellite are equipped with edge-servers providing marine computing services. The computation workloads of MASSs can be offloaded to UAVs and the LEO satellite in parallel via a multi-access approach. Then, we formulate an optimization problem and propose energy-efficient algorithms to minimize the energy consumption of SAMIN under latency constraints. Specifically, we exploit an alternating optimization (AO) method and a layered approach to decompose the original problem into four optimization problems (i.e., offloading mode optimization, offloading volume optimization, resource allocation of UAVs, and resource allocation of the LEO satellite) to obtain the optimal solutions. 
Numerical results are provided to verify the efficiency and effectiveness of the proposed scheme. 
\textcolor{b}{For future work, we plan to leverage artificial intelligence (AI) techniques, such as reinforcement learning or predictive analytics, to dynamically optimize resource allocation and task scheduling in response to varying environmental conditions and system demands. Additionally, we will incorporate advanced mobility models (e.g., stochastic mobility models) for MASSs to describe their movement patterns, for MASS collaborative trajectory planning and optimization.}
	
%These extensions will address the challenges posed by marine device mobility and differentiated marine service requirements, further improving the adaptability and scalability of the framework for real-world applications.

\section*{Acknowledgment}
This work was supported by the National Natural Science Foundation of China under Grant 62371085 and Grant 51939001.

\ifCLASSOPTIONcaptionsoff
  \newpage
\fi

\bibliographystyle{ieeetr}
\bibliography{Reference}

%\vspace{-10 mm}

\end{document}